\documentclass[onecolumn]{aastex631}

\usepackage{mathtools}
\usepackage{gensymb}
\usepackage[version=4]{mhchem}
\usepackage{soul}

\newcommand{\pyratbay}{\textsc{Pyrat Bay}}

\shorttitle{Kinetic and photochemical disequilibrium in WASP-69b}
\shortauthors{Bangera et al.}

\graphicspath{{./}{figures/}}
\begin{document}

\title{Kinetic and photochemical disequilibrium in the potentially carbon-rich atmosphere of the warm-Jupiter WASP-69b}

\author[0009-0008-7545-5022]{Nidhi Bangera}
\affiliation{Space Research Institute, Austrian Academy of Sciences, Schmiedlstrasse 6, 8042 Graz, Austria}
\affiliation{Institute for Theoretical and Computation Physics, Graz University of Technology, Petersgasse 16, 8010 Graz, Austria}

\author[0000-0002-8275-1371]{Christiane Helling}
\affiliation{Space Research Institute, Austrian Academy of Sciences, Schmiedlstrasse 6, 8042 Graz, Austria}
\affiliation{Institute for Theoretical and Computation Physics, Graz University of Technology, Petersgasse 16, 8010 Graz, Austria}

\author[0000-0002-1259-2678]{Gloria Guilluy}
\affiliation{INAF – Osservatorio Astrofisico di Torino, Via Osservatorio 20, 10025, Pino Torinese, Italy}

\author[0000-0002-1347-2600]{Patricio Cubillos}
\affiliation{INAF – Osservatorio Astrofisico di Torino, Via Osservatorio 20, 10025, Pino Torinese, Italy}
\affiliation{Space Research Institute, Austrian Academy of Sciences, Schmiedlstrasse 6, 8042 Graz, Austria}

\author[0000-0003-4426-9530]{Luca Fossati}
\affiliation{Space Research Institute, Austrian Academy of Sciences, Schmiedlstrasse 6, 8042 Graz, Austria}

\author[0000-0001-7034-7024]{Paolo Giacobbe}
\affiliation{INAF – Osservatorio Astrofisico di Torino, Via Osservatorio 20, 10025, Pino Torinese, Italy}

\author[0000-0002-7180-081X]{Paul Rimmer}
\affiliation{Cavendish Laboratory, University of Cambridge, JJ Thomson Ave, Cambridge, CB3 0HE, United Kingdom}

\author[0000-0003-4269-3311]{Daniel Kitzmann}
\affiliation{Space Research and Planetary Sciences, Physics Institute, University of Bern, Gesellschaftsstrasse 6, 3012 Bern, Switzerland}
\affiliation{Center for Space and Habitability, University of Bern, Gesellschaftsstrasse 6, 3012 Bern, Switzerland}

\begin{abstract}

High-resolution transmission spectroscopy of the warm gas-giant WASP-69b has revealed the presence of H$_2$O, CO, CH$_4$, NH$_3$, and C$_2$H$_2$ in its atmosphere. This study investigates the impact of vertical diffusion and photochemistry on its atmospheric composition, with a focus on the detected species plus HCN and CO$_2$, to constrain the atmospheric C/O ratio. We utilize non-equilibrium gas-phase simulations to conduct a parameter study for vertical diffusion strength, local gas temperature, and C/O ratio. Our results indicate that a carbon-rich atmosphere enhances CH$_4$ and C$_2$H$_2$ concentrations, while NH$_3$ undergoes chemical conversion into HCN in carbon-rich, high-temperature environments. Consequently, HCN is abundantly produced in such atmospheres, though its strong spectral features remain undetected in WASP-69b. Photochemical production of HCN and C$_2$H$_2$ is highly sensitive to vertical diffusion strength, with weaker diffusion resulting in higher concentrations. Cross-correlation of synthetic spectra with observed data shows that models with C/O=2 best match observations, but models with C/O=0.55 and 0.9 lead to statistically equivalent fits, leaving the C/O ratio unconstrained. We highlight the importance of accurately modeling NH$_3$ quenching at pressures greater than 100 bars. Models for WASP-69b capped at 100 bars bias cross-correlation fits towards carbon-rich values. We suggest that if the atmosphere of WASP-69b is indeed carbon-rich with a solar metallicity, future observations should reveal the presence of HCN. 

\end{abstract}


\section{Introduction} \label{sec:intro}

The atmospheric composition of exoplanets can offer valuable insights on planet formation processes. Previous studies suggest that atmospheric elemental ratios, such as the carbon-to-oxygen ratio (C/O), can help identifying the location of gas giant exoplanets' formation and their accretion history within a protoplanetary disk \citep[e.g.][]{Oberg2011,Helling2014,Oberg2016,Molliere2022,Tabone2023}. Whether the planetary C/O coincides with that of its host star could depend on the method of planet formation, with two possible scenarios: the ``inheritance'' scenario, where the planet inherits its chemical makeup from the parent molecular cloud, or  the ``chemical reset'' scenario, where the accreted material is reset to its atomic state due to the ionising irradiation from the protostar \citep{Eistrup2016, Eistrup2018}.  

Previously, there have been a few observations that indicate super-solar C/O in exoplanet atmospheres \citep[e.g.][]{Madhusudhan2011,Oppenheimer2013,Tsiaras2016}. One example is WASP-69b, a transiting warm-giant exoplanet with an equilibrium temperature of T$_{\rm eq} \sim$ 950K\citep{Anderson2014,Bonomo2017}. In their study, \citet{Guilluy} utilized high-resolution transmission spectroscopy to observe the atmosphere of WASP-69b, detecting H$_2$O, CH$_4$, NH$_3$, CO, and C$_2$H$_2$. \citet{Guilluy} employed radiative and thermochemical equilibrium models, incorporating disequilibrium-chemistry effects in a post-processing step, to explore different C/O ratios and atmospheric metallicity scenarios consistent with their observations. Their findings suggest that atmospheric scenarios with super-solar metallicities are unlikely for this planet. Additionally, they propose that the observed presence of C$_2$H$_2$ and the stronger detection of CH$_4$ compared to H$_2$O could be better explained by an atmospheric C/O ratio greater than 1. Among the models tested, the most favorable match to observations was achieved by a model with solar metallicity, C/O = 2.0, and artificially increased concentrations for NH$_3$ and C$_2$H$_2$ at 10$^{-7}$–$10^{-6}$. However, the high concentrations of NH$_3$ and C$_2$H$_2$ may also be explained by chemical disequilibrium processes in the atmosphere. Therefore, disequilibrium chemistry calculations are needed to confirm or disprove the conclusions of \citet{Guilluy} on the presence of disequilibrium chemistry and on the inferred C/O.  In this work, we build upon the findings of \citet{Guilluy} by employing a novel approach that couples kinetic modeling, which provides physically supported abundance profiles accounting for disequilibrium chemistry, and high-resolution transmission spectroscopy to constrain the chemical properties of the atmosphere.

At atmospheric pressures that can be probed by infrared observations (10$^{-6}$ -- 1 bar), vertical diffusion-induced quenching and photochemistry are the two key physical processes that may drive the atmospheric composition of an exoplanet out of thermochemical equilibrium.  In the deep atmosphere, where gas temperatures and pressures are high, the chemistry can be assumed to be in thermochemical equilibrium. However, at lower atmospheric gas temperatures, when chemical timescales are larger than dynamical timescales, vertical diffusion can push the chemistry out of equilibrium \citep[e.g.][]{Moses2014, Baxter2021, Kawashima2021}. Molecules may thus be present in the upper atmospheric layers with concentrations higher than in equilibrium, suggesting that their concentrations are ``quenched'' in the deep atmosphere at thermochemical equilibrium concentrations and then transported upwards in the atmosphere faster than they can be chemically converted to the expected equilibrium species higher up in the atmosphere. These processes are expected to be particularly significant for planets with T$_{\rm eq}$$\lesssim$1300 K \citep[e.g.][]{Madhusudhan2014, Moses2014} such as WASP-69b. At these temperatures, where the carbon chemistry shifts between being CH$_4$- and CO-dominated, while the nitrogen chemistry shifts between being NH$_3$- and N$_2$-dominated \citep{Lodgers2002,Woitke2018}, vertical diffusion induced quenching can significantly impact both the diversity of the observed species and the strength of the observed spectral features. \citet{Zamyatina2023} find that while this diffusion-induced quenching is predicted to occur in the atmospheres of HAT-P-11b, HD 189733b, HD 209458b, and WASP-17b, the extent to which it affects spectra and phase curves varies from planet to planet. 

Additionally, photochemistry, i.e. the impact of the stellar radiation on the gas composition, can influence atmospheric composition \citep[e.g.][]{Madhusudhan2012,Moses2013,Moses2014,Barth2021,Baeyens2022}. The parent molecules transported from deep atmospheric regions are photodissociated in upper layers producing radicals that react to form new species. The shape of the stellar spectra, as well as the atmospheric temperature profile and initial atmospheric elemental composition, influence the diversity of chemical products in the atmosphere. Lower atmospheric temperatures produce abundant CH$_4$ and NH$_3$ that is beneficial for photochemistry, since these species are more active photochemical precursors than CO and N$_2$ \citep{Moses2014}. High atmospheric temperatures serve to counter balance the effect of photochemistry, maintaining the atmospheric composition closer to what is expected in the presence of thermochemical equilibrium.  Recently, direct evidence of photochemistry in hot gas-giant atmospheres was obtained from the JWST observations of SO$_2$ in the atmospheres of WASP-39b \citep{Tsai2023} and WASP-107b \citep{Dyrek2023}. On WASP-69b, it is the detection of C$_2$H$_2$ that suggests that a photochemical model is preferable to chemical equilibrium models for interpreting the observations \cite[e.g.][]{Moses2011,Venot2012,Moses2014}. 

To investigate the impact of disequilibrium chemistry on the atmospheric composition of WASP-69b, we performed a series of 1D calculations using the ARGO code \citep{Rimmer&Helling}, which incorporates thermochemical and photochemical kinetics along with vertical transport processes to model vertical atmospheric profiles. The code currently utilizes the STAND2020 network, which includes neutral and ion chemistry for molecules containing H, C, O, and N, with a limited neutral chemistry for S-bearing molecules. Our study focuses on three key parameters: atmospheric C/O ratio, thermal structure, and the strength of eddy diffusion. The atmospheric C/O ratio is of particular interest, as equilibrium modeling has suggested a carbon-rich atmosphere for WASP-69b, which contrasts with the C/O ratio of its host star \citep{Anderson2014}. Additionally, we accounted for the temperature difference of up to $\sim$400 K between the two terminator regions, as expected for a planet with an equilibrium temperature of $\sim$1000 K orbiting a K-type star like WASP-69b \citep{Helling2023}. The eddy diffusion coefficient was also examined, given its role as a poorly constrained parameter that can significantly influence atmospheric concentration predictions, potentially varying by orders of magnitude \cite[e.g.][]{Moses2011,Arfaux2023}. Finally, synthetic spectra generated from our models were compared to the observations of \cite{Guilluy} to evaluate the consistency of our findings.

This paper is structured as follows. Section \ref{sec:Approach} describes the modeling approach  used to derive the atmospheric composition and synthetic spectra. Section \ref{sec:Results} presents how the disequilibrium results differ from the chemical equilibrium solution for our base model, and the results of the parameter space study for each of the detected species (H$_2$O, CO, CH$_4$, C$_2$H$_2$, NH$_3$), plus HCN and CO$_2$ that were not detected in the GIANO-B near-infrared data. Section \ref{sec:fitting} presents the synthetic transmission spectra in comparison to the observations of \cite{Guilluy}, together with the implications for WASP-69b's atmospheric C/O. Section \ref{sec:discussion} presents a discussion of the degeneracy in the models as well as the model sensitivities to boundary conditions. Section \ref{sec:conclusions} concludes the paper. 

\section{Modelling approach} 
\label{sec:Approach}

We employ STAND2020 \citep{Hobbs2021,Rimmer2021}, an ion-neutral photochemical C/N/O/H+S rate-network with $>$4000 reaction paths to determine the local gas-phase composition in the observable terminator regions of WASP-69b. Leaving the required input radiation field constant, we explore the effect of local temperature, mixing efficiency, and C/O on the atmospheric chemistry leading to 14 independent models described in Table \ref{parameters}. The chemical models are used as input to calculate transmission spectra, which are then cross-correlated with the near-IR observations to test the conclusions of \citet{Guilluy} employing a physically motivated model.

\subsection{Chemical Kinetics Modelling} \label{ARGO+STAND}
We employed the 1D photochemical diffusion code ARGO \citep{Rimmer&Helling,ErratumRimmer_2019} to model the atmospheric chemistry of WASP-69b in thermochemical disequilibrium. The model consists of two parts: a transport model that considers ion-neutral and neutral-neutral reactions, and a model calculating the chemical rate constants for photochemistry and cosmic rays. It solves the 1D continuity equation for each gas species $i$,
\begin{equation}
    \frac{\partial n_i}{\partial t} = P_i - L_i - \frac{\partial \phi_i}{\partial z},
\end{equation}
\noindent
where $n_i$, in cm$^{-3}$, is the number density of species $i$, $P_i$ and $L_i$ are the production and loss rates of species $i$, respectively [cm$^{-3}$s$^{-1}$], and $\frac{\partial \phi_i}{\partial z}$ is the vertical change in flux $\phi_i$ [cm$^{-2}$s$^{-1}$]. The production and loss terms $P_i$ and $L_i$ describe two-body neutral-neutral and ion-neutral reactions, three-body neutral reactions, dissociation reactions, radiative-association reactions, thermal ionization and recombination reactions.
Vertical transport is assumed to occur via both eddy ($K_{\rm{zz}}$ [cm$^{2}$s$^{-1}$]) and molecular ($D_{\rm{zz}}$ [cm$^{2}$s$^{-1}$]) diffusion,
\begin{equation}
    \phi_i = - K_{\rm{zz}} \left[ \frac{\partial n_i}{\partial z} + n_i \left( \frac{1}{H_0} + \frac{1}{T}\frac{dT}{dz} \right) \right] - D_{\rm{zz}}\left[ \frac{\partial n_i}{\partial z} + n_i \left( \frac{1}{H_i} + \frac{1 + \alpha_T}{T}\frac{dT}{dz} \right) \right],
\end{equation}
\noindent
where $H_0$ is the atmospheric pressure scale height at vertical height $z$, $H_i$ is the atmospheric scale height for species $i$, $T$ is the temperature, and $\alpha_T$ the thermal diffusion coefficient \citep{Banks1973,Zahnle2006}. The molecular diffusion coefficients in ARGO are adopted from Chapman-Enskog theory \citep{Enskog1917,Chapman1991}. The eddy diffusion coefficients are used as a free parameter. 
%
   \begin{table}
      \caption{WASP-69b system parameters for the host star and the planet as used by \citet{Guilluy}.}
      \centering 
      \def\arraystretch{1.3}
         \label{Input}
         \begin{tabular}{c|c|c}
             \hline
            Quantity      &  Value & Source\\
           \hline
            Stellar Mass  [M$_\odot$] & 0.826(29) & \cite{Anderson2014}\\
            Stellar Radius [R$_\odot$]    & 0.813(28) & \cite{Anderson2014}\\
            Stellar Effective Temperature [K] & 4715(50)  & \cite{Anderson2014}\\
            Planetary Mass  [M$_{\rm{Jup}}$]& 0.260(17) & \cite{Anderson2014} \\
            Planetary Radius [R$_{\rm{Jup}}$]    & 1.057(47) & \cite{Anderson2014} \\
            Planetary Equilibrium Temperature [K] & 963(18) &    \cite{Anderson2014}\\
            Semi-major axis [au] & 0.04527$^{+0.00053}_{-0.00054}$  &   \cite{Bonomo2017}\\
            \hline 
         \end{tabular}
      \tablecomments{The number in parentheses are the standard uncertainty corresponding to the last digits of the quoted values.}      
         
   \end{table}
%
\subsection{Radiative Transfer with C-DISORT}
In order to calculate photochemical rates, the stellar high-energy flux received at the top of the planet's atmosphere and its variation throughout the atmosphere is required. For the corresponding radiative transfer calculations, we added the discrete ordinate radiative transfer code C-DISORT \citep{Hamre2013AIPC.1531..923H} to ARGO. C-DISORT is an improved C version of the well-known DISORT code originally written in FORTRAN \citep{Stamnes1988ApOpt}.

C-DISORT provides the exact solution of the plane-parallel radiative transfer equation for a given set of polar angles. In the framework of a discrete ordinate radiative transfer schemes, these are referred to as streams and their number determines the numerical accuracy of the solution. For this work we employ four streams, which is usually sufficient for calculating angular-averaged quantities in scenarios without strongly asymmetric scattering. The C-DISORT radiative transfer yields the mean intensity as a function of atmospheric height which is subsequently converted into the actinic flux that is used to calculate the corresponding photodissociation rates.

\subsection{Model Setup}

For the incident stellar radiation, we constructed the stellar spectral energy distribution for WASP-69b (Figure~\ref{FigGam}) by mixing a PHOENIX model in the UV, optical, and infrared \citep{Husser2013} with a scaled solar flux in the EUV and X-ray regimes \citep{Claire2012}. The solar XUV emission has been scaled to best match the observed stellar X-ray luminosity \citep{Nortmann2018} and expected EUV flux \citep{Fossati2023HeI}. To model the terminator regions, we set the stellar zenith angle of the incident stellar beam in C-DISORT to 85$\degree$. In addition to stellar radiation, we include the effect of low-energy cosmic rays in our simulations based on \cite{Rimmer&Helling}. For atmospheric composition and synthetic transmission spectra calculations, we considered the same stellar parameters and planetary log(g) used by \citet{Guilluy} (Table \ref{Input}). 

\begin{figure*}
   \centering
   \includegraphics[width = 0.4\linewidth]{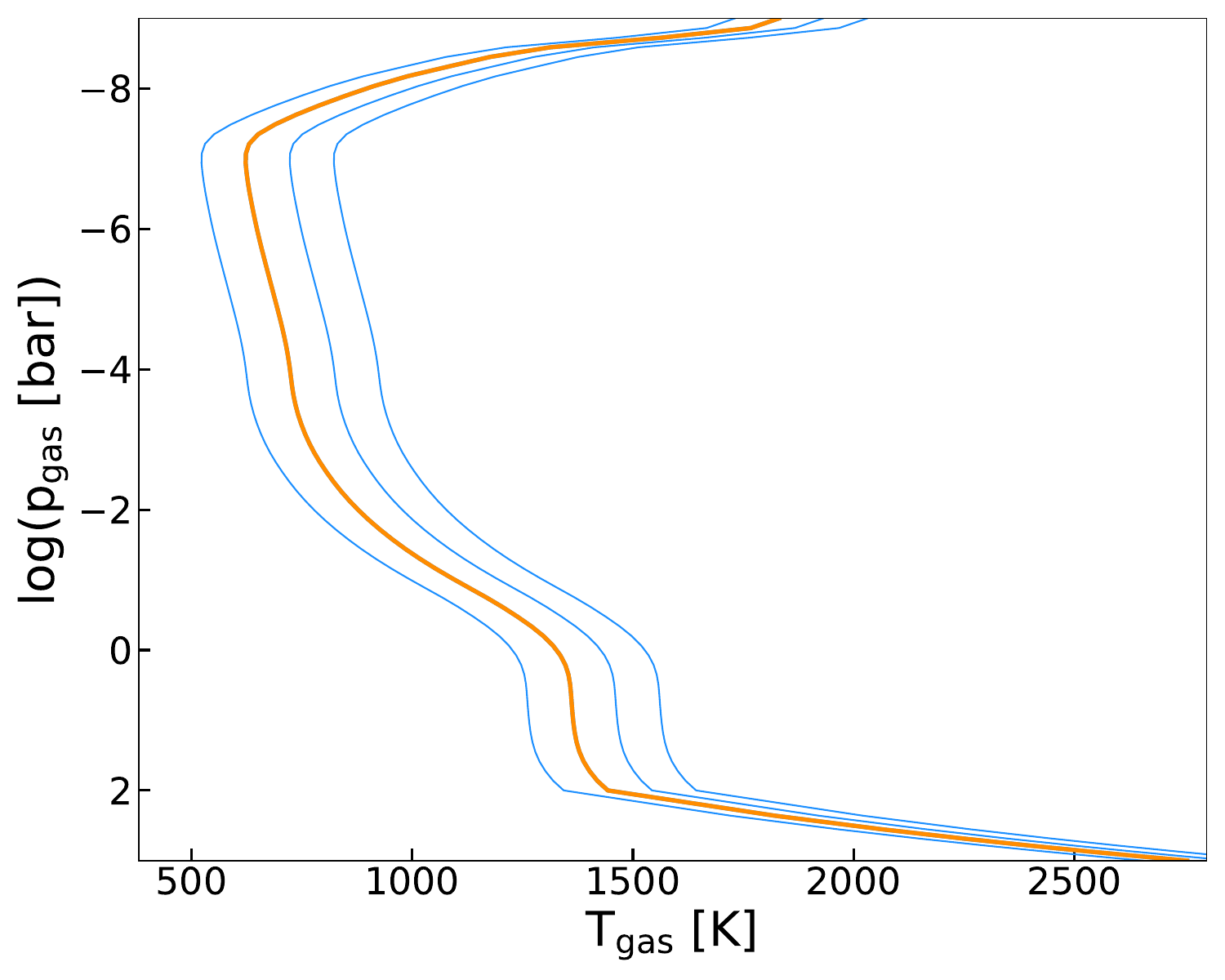}
   \includegraphics[width = 0.4\linewidth]{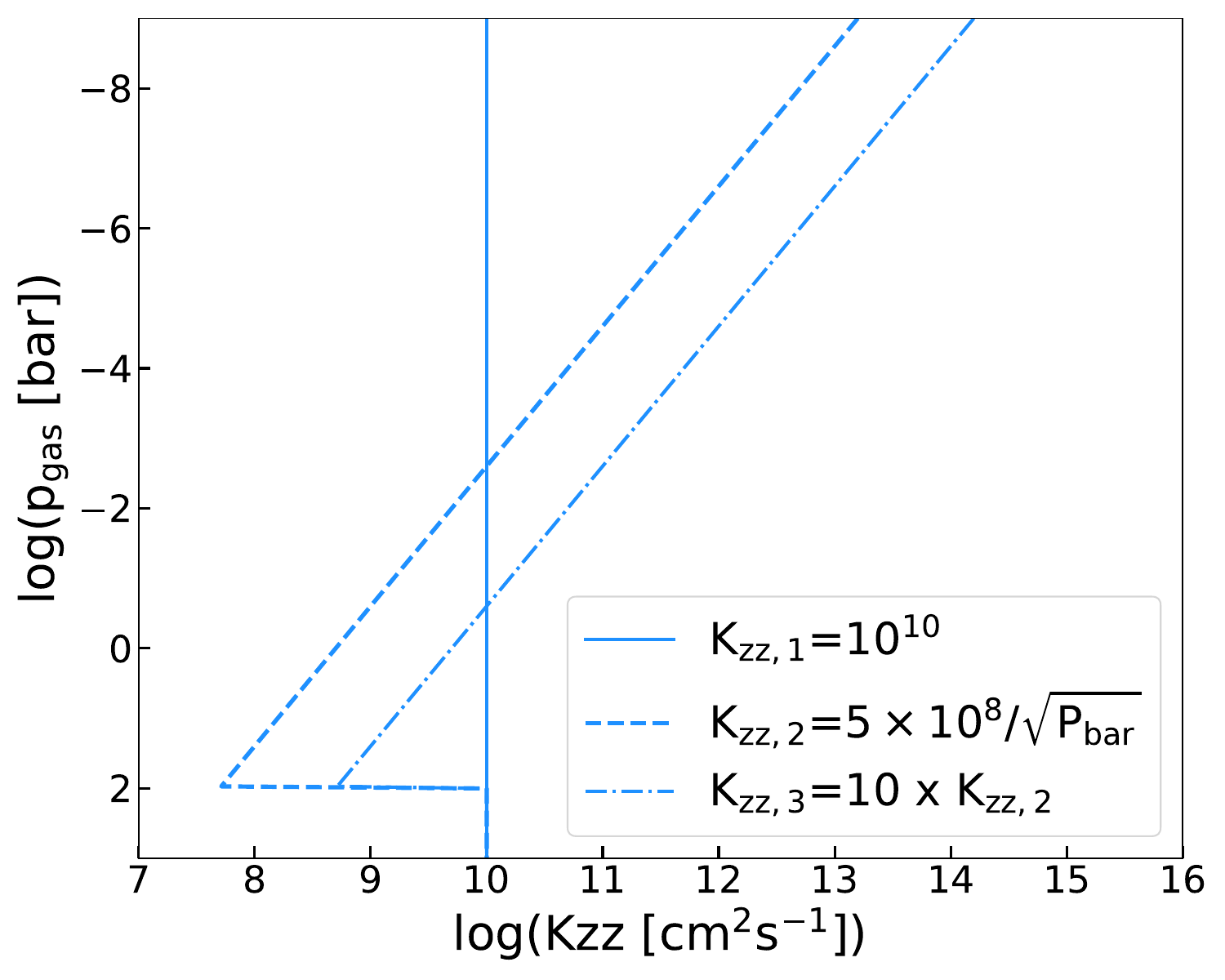}
   \caption{Gas temperature-pressure (left) and eddy diffusion K$_{\rm{zz}}$ (right) profiles used in this work. The orange (T$_{\rm gas}$, p$_{\rm gas}$) profile , upto 100 bars, is adapted from \cite{Guilluy} for their analysis of WASP-69b and extended to 1000 bars assuming a hot adiabatic temperature profile.} All other (T$_{\rm gas}$, p$_{\rm gas}$) profiles are obtained by shifting the orange profile in steps of 100 K. 
              \label{PT}
    \end{figure*}

We treated the local thermodynamic properties, element abundances, and mixing efficiency as free parameters in our models as described below:

1) \textbf{Temperature profile:} Figure \ref{PT} illustrates the (T$_{\rm gas}$, p$_{\rm gas}$) profiles used as input to model WASP-69b's atmosphere chemistry. The base profile (orange curve) was obtained from 1D radiative-thermochemical-equilibrium calculations assuming solar composition \citep[][Cubillos et al. 2024, in prep]{Guilluy}. These radiative-equilibrium calculation considered an atmospheric model ranging from 100 to 10$^{-9}$ bar, and a spectral window ranging from 0.3 to 30 $\mu$m to cover the bulk of the stellar and planetary radiation. The radiative transfer was computed with the {\pyratbay} code \citep{Cubillos2021}, using the line-by-line sampling approach over a cross-section grid with a resolving power of 15000.  The radiative-transfer routine in {\pyratbay} uses the line-sampling approach, considering the major sources of opacity expected in H$_2$-dominated atmospheres: CO, CO$_2$, SO$_2$, and CH$_4$ from HITEMP \citep{Rothman2010, Li2015, Hargreaves2020}; H$_2$O, HCN, NH$_3$, and C$_2$H$_2$ from ExoMol \citep{AzzamEtal2016mnrasExoMolH2S, Polyansky2018, Chubb2020, YurchenkoEtal2011mnrasNH3opacities, HarrisEtal2006mnrasHCNlineList, HarrisEtal2008mnrasExomolHCN, UnderwoodEtal2016mnrasSO2exoamesExomol, ColesEtal2019mnrasNH3coyuteExomol}; Na and K  \citep{BurrowsEtal2000apjBDspectra}; and collision-induced absorption (CIA) due to H$_2$--H$_2$ and H$_2$--He \citep{BorysowEtal1988apjH2HeRT,BorysowEtal1989apjH2HeRVRT, BorysowEtal2001jqsrtH2H2highT,BorysowFrommhold1989apjH2HeOvertones, Borysow2002jqsrtH2H2lowT}. The large molecular line-list databases were pre-processed with the \textsc{repack} algorithm
\citep{Cubillos2017apjRepack} before sampling into the opacity data. An intrinsic heat flux of 100~K was adopted, as it is typically assumed for Jupiter-sized planets.

To extend the pressure profile to 1000 bar, capturing deeper regions where NH$_3$ quenching may occur \citep{Moses2011}, we applied the hot adiabatic temperature profile from \citet{SainsburyMartinez2019} as detailed in equation 21 of \citet{Schneider2022},
\begin{equation}
    T(p>100 \rm{bar}) = \Theta_{ad} \cdot \left(\frac{p_{\rm{gas}}}{100 \rm{bar}}\right)^\eta,
\end{equation}
\noindent
where $\Theta_{ad}$ is the local gas temperature of the adiabat at 100 bars and $\eta$ = 1/3.56. This extended profile serves as our base model ``TP'', while additional profiles are obtained by shifting TP by $\pm$100 K throughout the atmosphere. 

2) \textbf{Eddy diffusion coefficient (K$_{zz}$):} While several studies using 3D general circulation models (GCMs) have provided numerical and theoretical estimates of the eddy diffusion coefficient K$_{\rm{zz}}$ \citep[e.g.][]{Parmentier,Komacek2019,Arfaux2023}, the parameter remains difficult to constrain. Three different K$_{\rm{zz}}$ profiles are investigated here (Figure \ref{PT}). The first is a constant K$_{\rm{zz,1}}$ = 10$^{10}$ cm$^{2}$s$^{-1}$, throughout the atmosphere. The second depends on the local gas pressure derived  by \citet{Parmentier} for HD\,209458b, $   K_{zz,2} = \frac{5 \times 10^{8}}{\sqrt{p_{gas}}} \rm{[cm^{2}s^{-1}]}\,.$
\citet{Parmentier} find this to be a valid parametrization for p$_{\rm gas}$=10$^{2}\,\ldots\,10^{-6}$ bar. We extend the K$_{\rm{zz, 2}}$ profile to 10$^{-9}$ bar, resulting in high K$_{\rm{zz}} (10^{13}-10^{14}$ cm$^2$s$^{-1}$) in the upper atmosphere. The K$_{\rm{zz,3}}$ profile is $10\times$ K$_{\rm{zz,2}}$, across the same pressure range. For pressures greater than 100 bar in all 3 profiles, we set Kzz to a constant value of 10$^{10}$ cm$^{2}$s$^{-1}$, as suggested by \citet{Moses2022}. The various  K$_{\rm{zz}}$ profiles investigate here allow to study a range of efficiencies in eddy mixing for both the top and bottom of the atmosphere.

3) \textbf{C/O:} The elemental abundances are set to solar \citep{Asplund2009}. For the composition of the atmosphere we consider C/O = 0.55 (solar), 0.9, 2 and 5, altering the C abundance each time. These values range from oxygen-rich to carbon-rich and are used to probe how high a value for C/O is required for C$_2$H$_2$ to be observable.

The base model utilizes the (T$_{\rm gas}$, p$_{\rm gas}$) profile ``TP'' (orange in Figure \ref{PT}), the pressure-dependent eddy diffusion profile K$_{\rm{zz,2}}\sim 1/{\sqrt{p_{gas}}} $, solar element abundances with an oxygen-rich C/O = 0.55. Thirteen additional models are built by independently varying these parameters. A summary of the different model parameters is presented in Table \ref{parameters}.

\begin{table}[ht!]
\caption{Description of Models Corresponding to Fig. A.1}             
\label{table:1}      
\centering                          
\begin{tabular}{ c | c | c | c }        
\hline\hline                 
Model Number & C/O & Temperature-Pressure profile (As in Fig. \ref{PT}) & K$_{\rm zz}$ (As in Fig. \ref{PT})\\    
\hline                        
    1 & 0.55 & TP & K$_{\rm{zz,1}}$ \\ 
    2 & 0.55 & TP    & K$_{\rm{zz,2}}$\\
    3 & 0.55 & TP     & K$_{\rm{zz,3}}$\\
    4 & 0.55 & TP-100    & K$_{\rm{zz,2}}$\\
    5 & 0.55 & TP+100    & K$_{\rm{zz,2}}$\\ 
    6 & 0.55 & TP+200 & K$_{\rm{zz,2}}$\\      
    7 & 2 & TP    & K$_{\rm{zz,1}}$\\
    8 & 2 & TP     & K$_{\rm{zz,2}}$\\
    9 & 2 & TP    & K$_{\rm{zz,3}}$\\
    10 & 2 & TP-100    & K$_{\rm{zz,2}}$\\ 
    11 & 2 & TP+100 & K$_{\rm{zz,2}}$\\      
    12 & 2 & TP+200    & K$_{\rm{zz,2}}$\\
    13 & 0.9 & TP     & K$_{\rm{zz,2}}$\\
    14 & 5 & TP    & K$_{\rm{zz},2}$\\
   
\hline                                   
\end{tabular}
\label{parameters}
\end{table}

\subsection{Cross-correlation to Observations}
The 14 non-equilibrium chemistry models computed in this work are compared to the ``full model'' used in \cite{Guilluy}, where the authors conducted high-resolution transmission spectroscopy of WASP-69b. They analyzed data from three nights gathered with the near-infrared spectrograph GIANO-B of the Telescopio Nazionale Galileo.

To compute the transmission spectra of WASP-69b based on our photochemical kinetics result, we used the {\pyratbay} package, this time sampling the cross sections at a resolution of R=80000 over a wavelength range of 0.9–2.5 µm, to be able to compare the results at a high spectral resolution.

To maintain consistency, we employed the same data processing methodology described in \cite{Guilluy} starting from the residual GIANO-B spectra resulting from the Principal Component Analysis (PCA) for telluric correction. The real data are cross-correlated with our models by following the same approach described in Section~3 of \citet{Guilluy}. Firstly, we convolved each model to the GIANO-B instrument profile (a Gaussian profile with full width at half maximum FWHM of $\sim$5.4 km s$^{-1}$). For each night, we used the GIANO-B orders selected in \citet{Guilluy}, and performed cross-correlation (CC) for every phase over a lag vector corresponding to planet radial velocities (RVs) in the range -252 $\leq$ RV $\leq$ 252 km s$^{-1}$, in steps of 3 km s$^{-1}$ (to cover all possible RVs of the planet). Subsequently, we co-added the CC functions over the selected orders, nights, and orbital phase after shifting them in the planet rest frame by assuming a circular orbit (i.e. zero eccentricity). We scanned a range of planet RV semi-amplitudes 0$\leq$K$_\mathrm{P}$$\leq$200 km s$^{-1}$, in steps of 1.5 km s$^{-1}$ (which includes the expected K$_\mathrm{P}$, K$_\mathrm{P,theo}$ = 127.11$^{+ 1.49 }_{- 1.52 }$ km/s, see Table~1 in \citealt{Guilluy}). To quantify the confidence level of our detections, we converted the total CC signal into signal-to-noise (S/N), by dividing the total CC matrix by its standard deviation (calculated by excluding the CC peak, e.g., \citealt[][]{Brogi_2018_Giano, Giacobbe2021}).

\section{Results} \label{sec:Results}

\subsection{Equilibrium vs Quenching vs Photochemistry} \label{sec:eqvsdis}

\begin{figure*}
     \centering
     \includegraphics[width=0.8\textwidth]{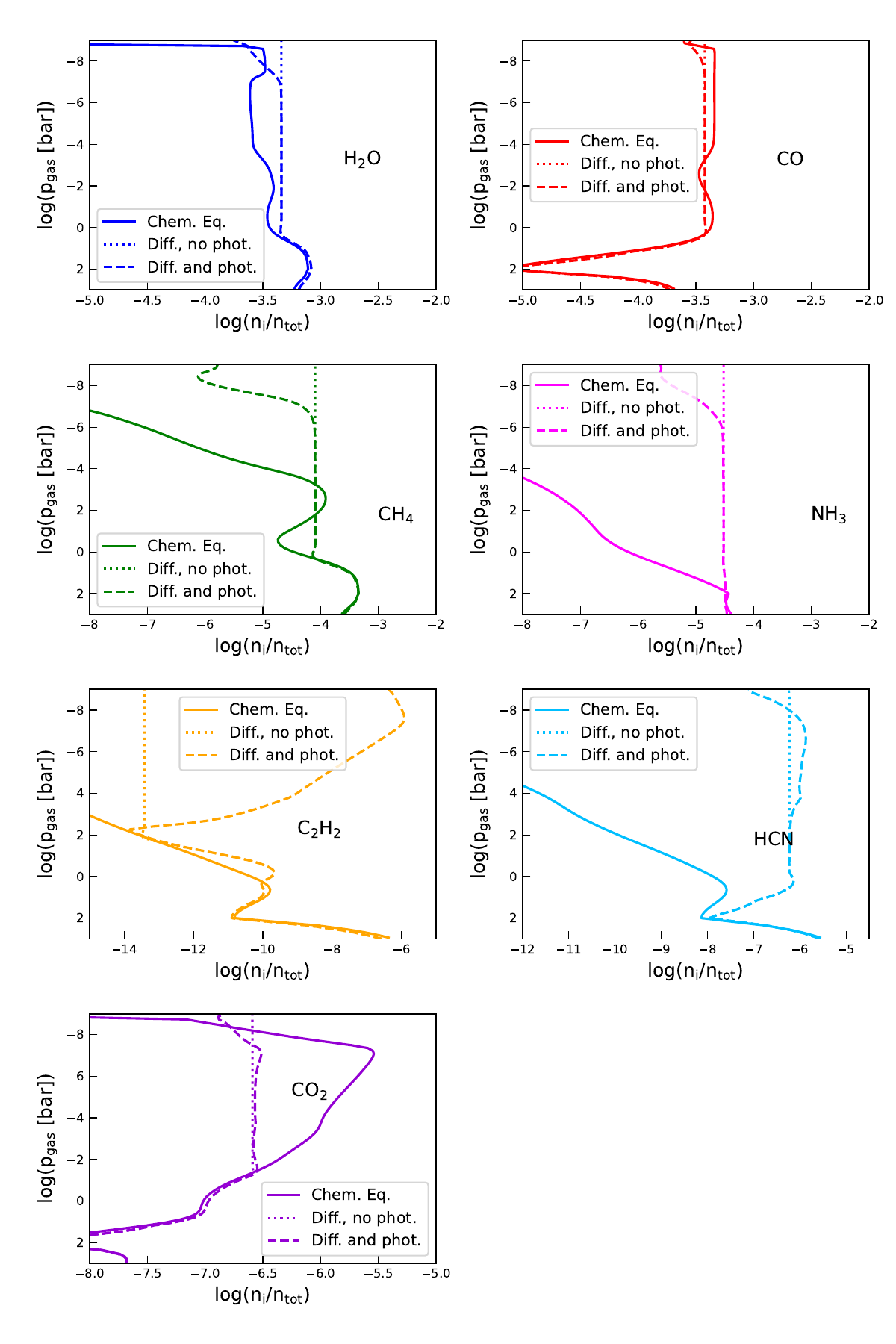}
        \caption{Local non-equilibrium gas-phase concentrations of H$_2$O, CO, CH$_4$, NH$_3$, C$_2$H$_2$, HCN, and CO$_2$ for i)  thermochemical equilibrium  (solid lines), ii) vertical diffusion but no photochemistry (dash-dot lines), and iii) vertical diffusion with photochemistry (dashed lines). The input (T$_{\rm gas}$, p$_{\rm gas}$) profile is the base model ``TP'' with the eddy diffusion profile K$_{\rm{zz,2}}$, and C/O=0.55. 
        NH$_3$ and CH$_4$ concentrations are increased by diffusion, which then contributes to increased concentrations of C$_2$H$_2$ and HCN. Note that the x-axis range differs in each subfigure.}
        \label{fig:three graphs}
\end{figure*}

The chemical species investigated in this work are H$_2$O, CO, CH$_4$, C$_2$H$_2$, NH$_3$, HCN, and CO$_2$. Of these, H$_2$O, CO, CH$_4$, C$_2$H$_2$ and NH$_3$ have been detected in the terminator regions of WASP-69b based on high-resolution transmission spectra collected with the GIANO-B near-infrared spectrograph. Ion-neutral chemistry has been included in our modeling with STAND, but its impact on the aforementioned molecules is found to be negligible.
We test the hypothesis by \citet{Guilluy} that these species can only be present simultaneously if chemical disequilibrium processes affect their formation and/or destruction. Since each of these molecules may be affected to a different extent by photochemistry and vertical diffusion in the atmosphere, three different cases are studied and compared: i) chemical equilibrium, ii) disequilibrium chemistry resulting from eddy diffusion, and iii) disequilibrium chemistry resulting from eddy diffusion and photochemistry.
The equilibrium concentrations (case i) are calculated using the GGchem code \citep{Woitke2018} using solar element abundances and C/O=0.55. We use the respective (T$_{\rm gas}$, p$_{\rm gas}$) profile as our base model as outlined in Sect.~\ref{ARGO+STAND} for cases ii) and iii) with  K$_{\rm zz, 2}$. The results are shown in Figure \ref{fig:three graphs}.

In both cases with ii) only diffusion and iii) diffusion with photochemistry, the concentrations of H$_2$O and CO remain close to their equilibrium values, except for p$_{\rm{gas}} < 10^{-7}$ bar, where they are depleted by photodissociation. This agreement with equilibrium concentrations is because both species are already the most abundant species at their quench points and throughout the atmosphere, and thus not significantly affected by quenching. This is consistent with the results presented in \cite{Moses2011}, \cite{Drummond2020}, and \cite{Baeyens2022}. 
CH$_4$, being less abundant in WASP-69b's atmosphere than CO at these gas temperatures, is more significantly affected by diffusion. It is quenched at $\sim$ 0.4 bar and photodissociated at p$_{\rm{gas}} < 10^{-6}$, resulting in concentrations up to 4 dex larger than its equilibrium concentrations in the upper atmosphere.  

NH$_3$ concentrations depart from equilibrium and are quenched at $\sim$ 100 bar. Similar high-pressure quenching for NH$_3$ has been reported by \cite{Moses2011} in the atmosphere of HD189733b, due to ammonia reaching the N$_2$-NH$_3$ interconversion quench point at high pressures and temperatures. We find that the quenching of NH$_3$ results in it having comparable concentrations to N$_2$ on WASP-69b. In our model, NH$_3$ is produced through the following scheme

\begin{equation}
\begin{aligned}
    \ce{N2 + H &-> N2H} \\
    \ce{N2H + H2 &-> N2H2 + H} \\
    \ce{N2H2 + H &-> NH2 + NH}\\
    \ce{NH + H2 &-> NH2 + H}\\
    \ce{2(NH2 + H2 &-> NH3 + H)}\\
    \ce{2H +M &-> H2 + M}\\
    \hline
    {\rm Net:~}\ce{N2 + 3H2 &-> 2NH3} \\
    \hline
    \label{N2-NH3}
\end{aligned}
\end{equation}
Here, M is any third body. The final quench point for NH$_3$ is at the intermediate gas pressures of $\sim$ 1 bar. This arises because the gas temperature profile remains relatively constant in the pressure range between 1-100 bar, allowing the effective interconversion reactions between NH$_3$ and N$_2$ to persist \citep{Moses2011}. At p$_{\rm gas}<1$ bar the temperature gradients become significant again (Fig \ref{PT}), causing the rates of the interconversion reactions to drop off, resulting in the final quenching of NH$_3$, beyond which it maintains a uniform concentration, until photochemical destruction in the upper atmospheric layers. The impact of this quench point is more pronounced when comparing models with different K$_{zz}$ values, as shown in Fig. \ref{C2H2_NH3}. 

C$_2$H$_2$ follows its equilibrium profile up to $\sim$ 1 bar. At 0.01 bar $< p_{\rm{gas}} <$ 1 bar, the disequilibrium concentrations are up to 1 dex higher than the equilibrium concentrations, caused by the increased concentrations of parent molecule CH$_4$ due to quenching. C$_2$H$_2$ production occurs through the following scheme at these pressure levels 
\begin{equation}
\begin{aligned}
    \ce{H + CH4 &-> CH3 + H2}\\
    \ce{CH3 + CH4 &-> C2H5 + H2} \\
    \ce{C2H5 + M &-> C2H4 + H + M} \\
    \ce{C2H4 + H &-> C2H3 + H2}\\
    \ce{C2H3 + M &-> C2H2 + H + M}\\
    \hline
    {\rm Net:~}\ce{2CH4 &-> C2H2 + 3H2} \\
    \hline
    \label{CH4-C2H2}
\end{aligned}
\end{equation}
At p$_{\rm{gas}}$ $\sim$ 0.01 bar, C$_2$H$_2$ is quenched. Its concentration is further enhanced by up to 7 dex in the upper atmosphere due to the abundance of photochemically released atomic H, along with high levels of quenched CH$_4$. The following chemical scheme is responsible for the production of C$_2$H$_2$ in the upper atmosphere  
\begin{equation}
\begin{aligned}
        \ce{2(CH4 + H &-> CH3 + H2)}\\
        \ce{2CH3 &-> C2H6}\\
        \ce{C2H6 + H &-> C2H5 + H2}\\
        \ce{C2H5 + H &-> C2H4 + H2}\\
        \ce{C2H4 + H &-> C2H3 + H2}\\
        \ce{C2H3 + H &-> C2H2 + H2}\\
        \ce{3H2 + M &-> 6H + M}\\
        \hline
        {\rm Net:~}\ce{2CH4 &-> C2H2 + 3 H2} \\
        \hline
        \label{CH4-C2H2 upper atmosphere}
\end{aligned}
\end{equation}
This scheme also results in the release of molecular hydrogen in the upper (photochemically active) atmosphere. It is similar to 
\cite{Moses2013} for the production of C$_2$H$_2$ in the atmosphere of XO-1b, with the exception of the step that produces C$_2$H$_4$. In their scheme, the production of C$_2$H$_4$ is accomplished through collision with a third body: \ce{C$_2$H$_5$ + M -> H + C$_2$H$_4$ + M}, resulting in the use of 2 fewer H and the production of 1 fewer H$_2$ in this reaction. 

Similar to C$_2$H$_2$, the concentration of HCN is influenced by chemical disequilibrium in three ways. First, in the deep atmosphere, the increased presence of parent molecules CH$_4$ and NH$_3$ leads to a higher chemical production of HCN. The production scheme is then as follows 
\begin{equation}
    \begin{aligned}
        \ce{NH3 + H &-> NH2 + H2}\\
        \ce{NH2 + H &-> NH + H2}\\
        \ce{NH + H &->  N + H2}\\
        \ce{CH4 + H &-> CH3 + H2}\\
        \ce{N + CH3 &-> H2CN + H}\\
        \ce{H2CN + M &-> HCN + H}\\
        \ce{H2 + M &-> 2H + M}\\
        \hline
        {\rm Net:~} \ce{NH3 + CH4 &-> HCN + 3H2}\\
        \hline
        \label{HCN formation}
    \end{aligned}
\end{equation}
Second, HCN concentrations are quenched at p$_{\rm{gas}}$ $\sim$ 1 bar. Third, XUV radiation boosts HCN at p$_{\rm{gas}}$ = 10$^{-3.5}$--10$^{-8}$ bar and photodissociates it at higher altitudes. At these gas pressures, the production scheme is similar to scheme \ref{HCN formation}, except for the formation of NH, which happens through the photodissociation of NH$_3$ through: \ce{NH3 ->[h$\nu$] NH + 2H}. In agreement with \cite{Drummond2020}, CO$_2$ is found to be depleted by a factor $\sim$4 at p$_{\rm gas}<10^{-1}$bar through quenching as carbon is quenched into CH$_4$. CO$_2$ is depleted by photodissociation at p$_{\rm{gas}}<10^{-7}$ bar. 

\subsection{The parameter study for individual molecules} 
 
\label{sec:parameter study}
\subsubsection{H2O}

\begin{figure}
\centering
\includegraphics[width = 0.48\linewidth]{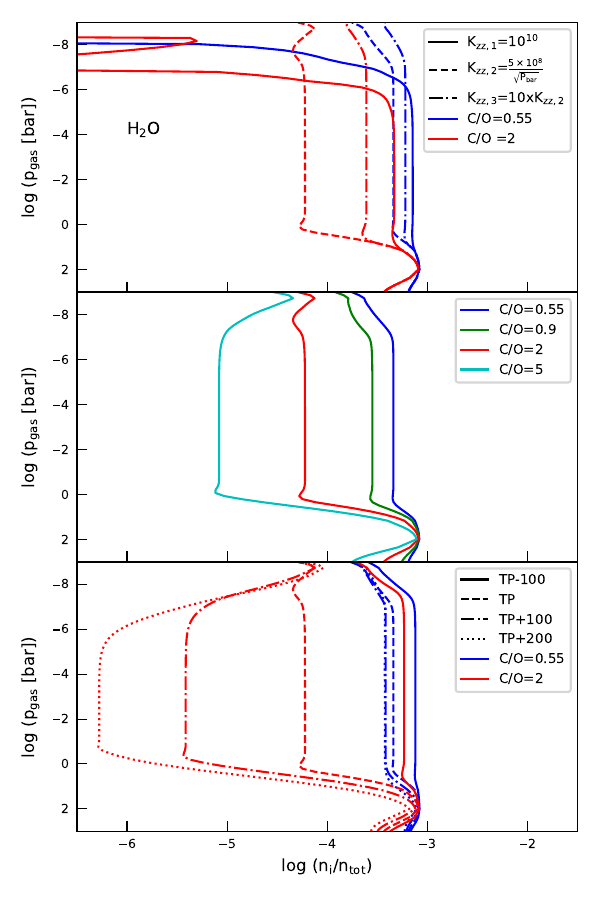}
\includegraphics[width = 0.48\linewidth]{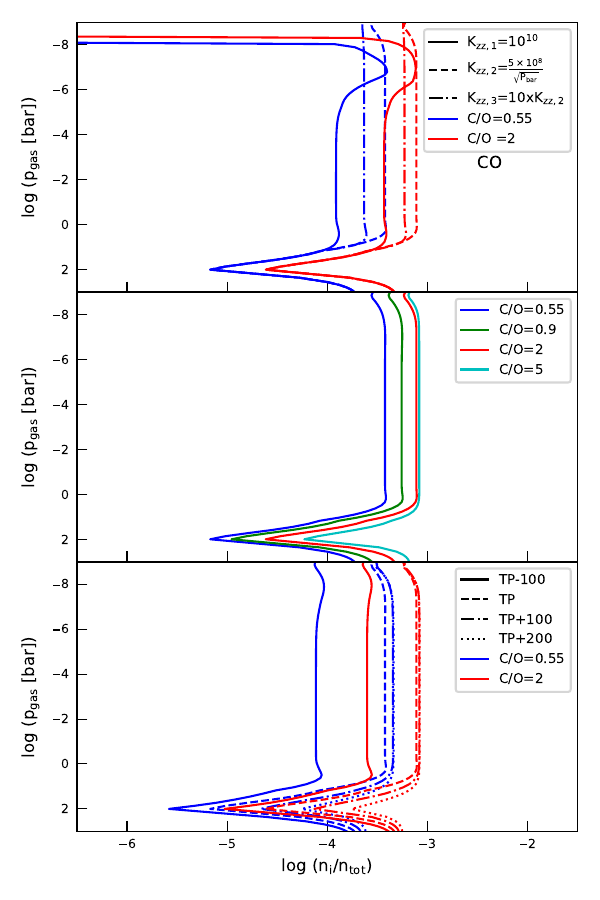}
\caption{H$_2$O and CO concentrations for the terminator region of WASP-69b. \textbf{Top:} varying eddy diffusion, for TP, C/O=0.55 and 2. \textbf{Middle:} varying C/O ratios, for TP and K$_{\rm{zz,2}}$. \textbf{Bottom:} varying (T$_{\rm gas}$, p$_{\rm gas}$), for K$_{\rm{zz,2}}$ and C/O=0.55 and 2.}
\label{H2O_CO}
\end{figure}

As shown in Figure \ref{H2O_CO}, at solar C/O the strength of the eddy diffusion parameter does not significantly affect H$_2$O concentrations in the deep atmosphere. However, in the upper atmosphere, the stronger eddy diffusion can support concentrations against photodissociation, whereas for the weakest eddy diffusion probed, H$_2$O concentrations are depleted by $> 4$ dex. At higher C/O, for e.g. C/O=2, H$_2$O is no longer the dominant species at its quench point and hence the strength of mixing significantly affects the concentrations. Here, the model with the most efficient mixing in the deep atmosphere (K$_{\rm{zz,1}}$) quenches at higher pressures and thus at higher concentrations. 
For ``TP'' and K$_{\rm{zz,2}}$, an increase in C/O results in decreased concentrations for H$_2$O as there is less oxygen available to form water once carbon sequesters the amount needed to form CO. This change in concentrations is less notable when C/O$<1$, i.e. in an oxygen dominated atmosphere.  
For C/O = 0.55, a change in temperature does not significantly affect H$_2$O concentrations. At colder temperatures there is a slight increase in concentration as CO depletes in the atmosphere in favour of CH$_4$. In contrast, for C/O = 2, an increase in gas temperature causes H$_2$O concentrations to deplete rapidly as more oxygen is preferentially trapped in CO. 

\subsubsection{CO}

At a planetary equilibrium temperature of $\sim$950 K, the CO--CH$_4$ interconversion becomes crucial in WASP-69b's atmosphere, with CO being the preferred species at higher temperatures. Thus, the strength of eddy diffusion does substantially affect the concentrations of CO, as seen in Figure \ref{H2O_CO}. For a higher diffusion coefficient, CH$_4$ is quenched at higher concentrations, and consequentially CO is quenched at a lower concentration. This trend remains consistent for both C/O of 0.55 and 2. In the upper atmosphere, a higher diffusion strength serves to replenish concentrations against photodissociation, similarly to H$_2$O.  
As C/O increases, the concentration of CO increases until oxygen is entirely trapped in CO. Further increase in C/O only results in higher concentrations of CH$_4$ and other hydrocarbons. 
With regards to the dependence on the local gas temperature, high temperatures push the carbon chemistry towards CO instead of CH$_4$. For WASP-69b this suggests that, at solar C/O, a gas temperature increase of $\sim$100K in the lower atmosphere can effectively turn the bulk of carbon from CH$_4$ into CO. The following chemical pathway converts CH$_4$ to CO in the upper atmosphere
\begin{equation}
    \begin{aligned}
        \ce{H2O &->[hv] H + H + O} \\
        \ce{CH4 + H &-> CH3 + H2} \\
        \ce{CH3 + O &-> CH2O + H} \\
        \ce{CH2O + H &-> CHO + H2}\\
        \ce{CHO + H &-> CO + H2} \\
        \hline
        {\rm Net:~} \ce{H2O + CH4 &-> CO + 3H2}\\
        \hline
        \label{CO-CH4}
    \end{aligned}
\end{equation}
\noindent

\subsubsection{CH4}

\begin{figure}
\centering
\includegraphics[width = 0.48\linewidth]{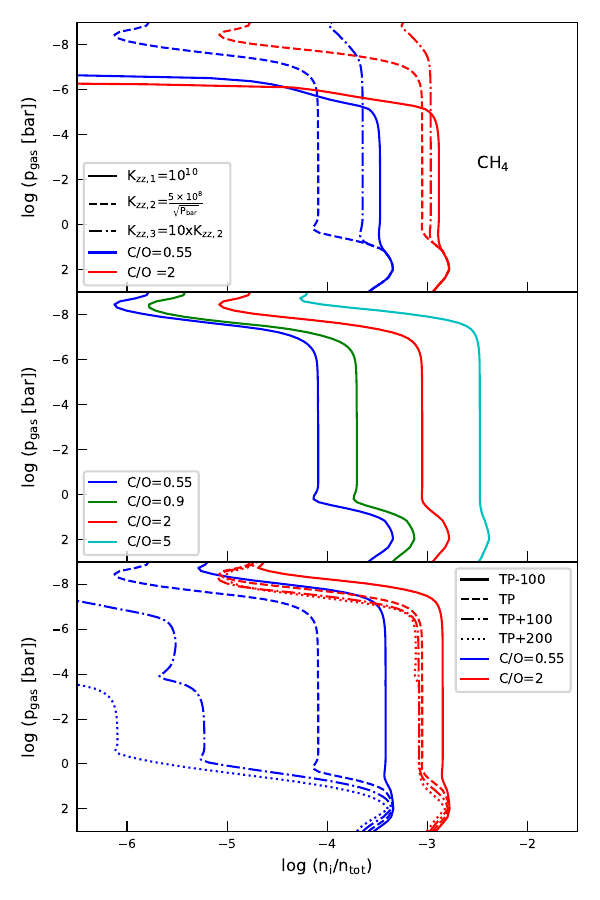}
\includegraphics[width = 0.48\linewidth]{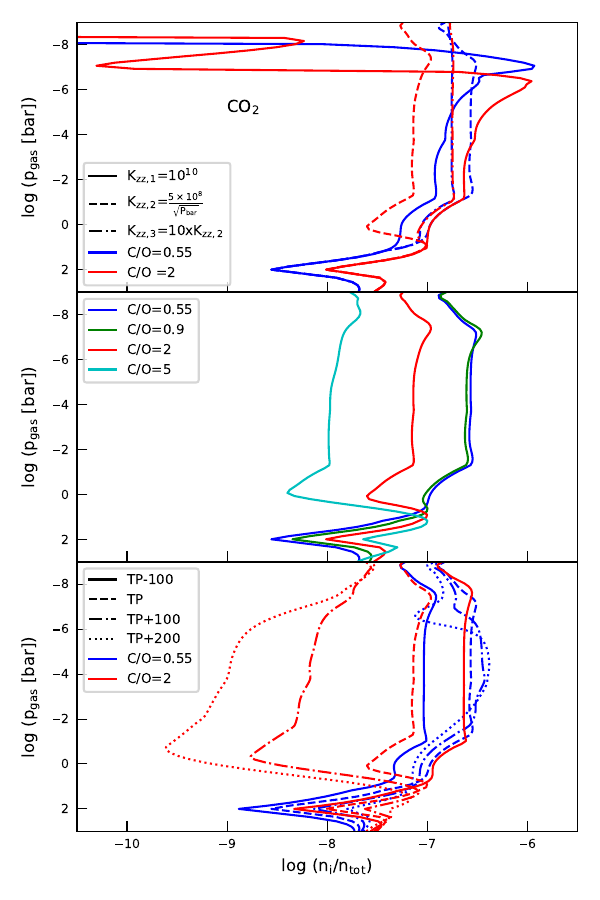}
\caption{Same as Figure \ref{H2O_CO} but for CH$_4$ and CO$_2$.}
\label{CH4_CO2}
\end{figure}

Figure \ref{CH4_CO2} shows the effect of the different parameters on CH$_4$ concentrations. Conversely to CO, a stronger mixing serves to increase the concentrations of CH$_4$ in the upper observable atmosphere. In the upper atmosphere, CH$_4$ is photodissociated, resulting in higher order hydrocarbons, and is also chemically recycled into CO through scheme \ref{CO-CH4}.
CH$_4$ concentrations experience a significant gain when C/O$>1$. As the equilibrium CH$_4$ concentration is larger at the quench point for the higher C/O models, the quenched methane concentration is larger for these models. This larger concentration then enhances the effectiveness of mechanisms that convert CH$_4$ to hydrocarbons like C$_2$H$_2$ (for e.g. through schemes \ref{CH4-C2H2} and \ref{CH4-C2H2 upper atmosphere}). 
For a solar C/O, there is a rapid depletion of CH$_4$ concentrations as the gas temperature increases, as the carbon chemistry shifts towards CO. However, for higher C/O, CH$_4$ remains the preferred carrier of carbon due to the limited amount of oxygen available in the atmosphere, and hence there is no strong depletion of concentrations with increase in temperature.

\subsubsection{CO2}

The concentration of CO$_2$ in the atmosphere of WASP-69b varies in response to the disequilibrium abundances of H$_2$O and CO via the following scheme
\begin{equation}
\begin{aligned}
    \ce{H2O + H &-> H2 + OH} \\
    \ce{CO + OH &-> CO2 + H} \\
    \hline
    {\rm Net:~} \ce{H2O + CO &-> H2 + CO2} \\
    \hline
    \label{CO formation}
\end{aligned}
\end{equation}
Thus, as seen in Fig \ref{CH4_CO2}, for a low C/O the reaction rate is limited by the concentration of the less abundant CO, and thus CO$_2$ reflects the same dependence on the eddy diffusion strength and local gas temperature as CO. Conversely, for a high C/O, H$_2$O concentrations are strongly depleted, and thus the dependence of CO$_2$ on K$_{\rm{zz}}$ and gas temperature are similar to that of H$_2$O. 
CO$_2$ is therefore also created more abundantly in an oxygen rich environment, as it needs to sequester the oxygen from H$_2$O. 

\subsubsection{C2H2}
\begin{figure}
\centering
\includegraphics[width = 0.48\linewidth]{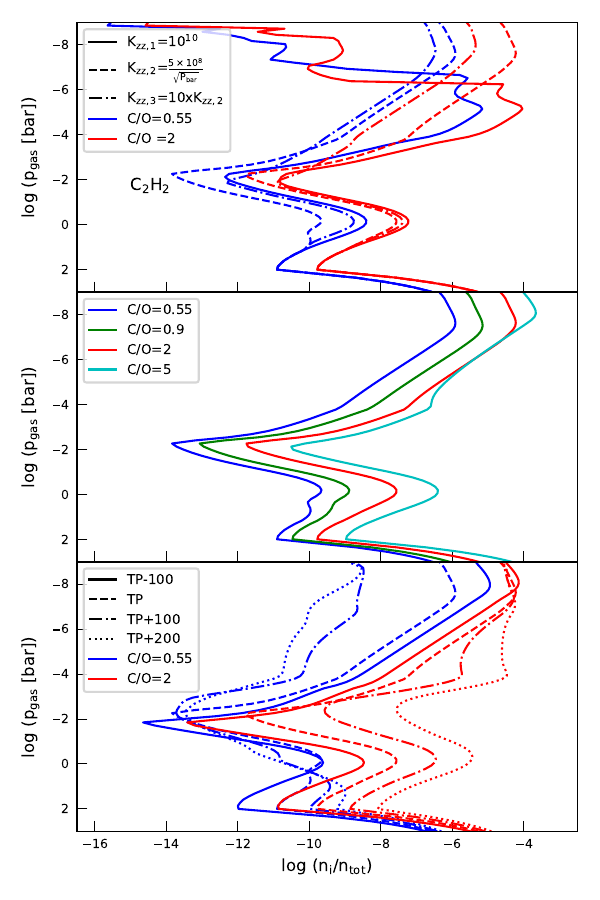}
\includegraphics[width = 0.48\linewidth]{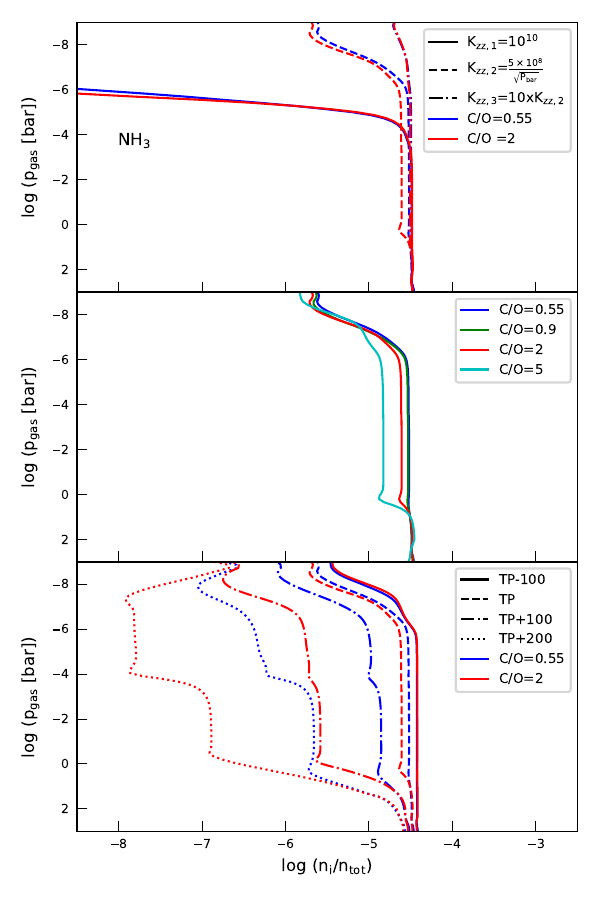}
\caption{Same as Figure \ref{H2O_CO} but for C$_2$H$_2$ and NH$_3$.}
\label{C2H2_NH3}
\end{figure}

The relationship between K$_{\rm{zz}}$ and C$_2$H$_2$ concentrations is complex, with a behavior that is not uniform throughout the atmosphere and depends on the local abundances of other species, as illustrated in Figure \ref{C2H2_NH3}. 
At higher pressures (10$^{-2}$--10$^{2}$ bars), C$_2$H$_2$ concentrations increase with stronger mixing (K$_{\rm{zz,1}}$), as C$_2$H$_2$ is formed through scheme \ref{CH4-C2H2} by the neutral-neutral reactions of CH$_4$ and CH$_3$, and thus reflects methane's dependency on K$_{\rm{zz}}$ at these pressures. Conversely, in the upper atmosphere (p$_{\rm{gas}} <10^{-2}$ bar), C$_2$H$_2$ concentrations decrease for higher mixing strengths, because C$_2$H$_2$ formation here requires the interaction of photochemically released H with the quenched CH$_4$. Stronger diffusion in the upper atmosphere works against photochemistry, leading to a lower production of C$_2$H$_2$.
In the lower atmosphere (p$_{\rm{gas}}> 10$ mbar), the dependence on K$_{\rm{zz}}$ is more prominent for the solar C/O model, as CH$_4$ concentrations depend strongly on the K$_{\rm{zz}}$. In the upper atmosphere, the dependence on K$_{\rm{zz}}$ is stronger for C/O=2, as C$_2$H$_2$ concentrations are impacted by the strong depletion of H$_2$O against the XUV flux for small K$_{\rm{zz}}$ values. When the C/O ratio exceeds 1, the concentration of C$_2$H$_2$ increases significantly due to the increased efficiency of the formation pathway through schemes \ref{CH4-C2H2} and \ref{CH4-C2H2 upper atmosphere}. This is in agreement to the results of \cite{Helling2017}, where the effect of increasing thermal stability of more complex hydrocarbon molecules with increasing C/O ratios was demonstrated for brown dwarfs.
Similarly to CO$_2$, there is an inversion in the dependence on temperature for low and high C/O. At low C/O, an increase in temperature leads to more CO production rather than CH$_4$, which in turn makes the production pathway of C$_2$H$_2$ through CH$_4$ less efficient. However, at high C/O, the concentration of CH$_4$ increases and the production pathway to form C$_2$H$_2$ becomes more efficient at higher temperatures, resulting in higher concentrations of C$_2$H$_2$. In all models, C$_2$H$_2$ is primarily lost into larger hydrocarbons such as C$_2$H$_4$ and C$_2$H$_6$, which have similar concentrations to C$_2$H$_2$. 
\cite{Gasman2022} suggest that a concentration above 1$\times$10$^{-7}$--1$\times$10$^{-6}$ is needed to observe hydrocarbon species through the atmosphere. This is also the requirement placed in \cite{Guilluy} for their best fit model. Our models show that these concentrations are achieved at high C/O, high temperatures, and low K$_{\rm{zz}}$. 

\subsubsection{NH3}

Figure \ref{C2H2_NH3} presents the effect of the different parameters on NH$_3$ concentrations. A more efficient atmospheric gas diffusion helps to maintain the quenched NH$_3$ concentration up to higher altitudes, as the fast eddy diffusion offsets the NH$_3$ loss caused by photodissociation (see also \citealt{Ohno2023}). NH$_3$ quenching occurs deep in the atmosphere at p$_{\rm{gas}}\sim 100$ bar, and since the K$_{\rm{zz}}$ values at the model lower boundary are identical across the profiles, no difference in NH$_3$ concentrations are observed at these depths. At pressures of $\sim$ 1 bar the second quench point emerges. Here the differences in K$_{\rm{zz}}$ between the profiles - approximately 1 dex - impacts NH$_3$ concentrations by a factor of 2.
Molecules such as NH$_3$ and N$_2$, that do not contain carbon or oxygen, are relatively unaffected by the C/O ratio. However, as the quenched concentration of CH$_4$ increases with higher C/O ratios, processes that kinetically convert CH$_4$ and NH$_3$ into HCN (scheme \ref{HCN formation}) become more effective for high C/O. This leads to a lower NH$_3$ concentration in the middle atmosphere above the quench point for the C/O = 2.0 model as compared to the C/O = 0.5 model. 
At these temperatures, nitrogen is preferentially stored in NH$_3$. For both a solar and C/O=2 model, NH$_3$ concentrations are depleted as temperatures increase and NH$_3$ is converted to HCN and N$_2$.  

\subsubsection{HCN}
\begin{figure}
\centering
\includegraphics[width = 0.48\linewidth]{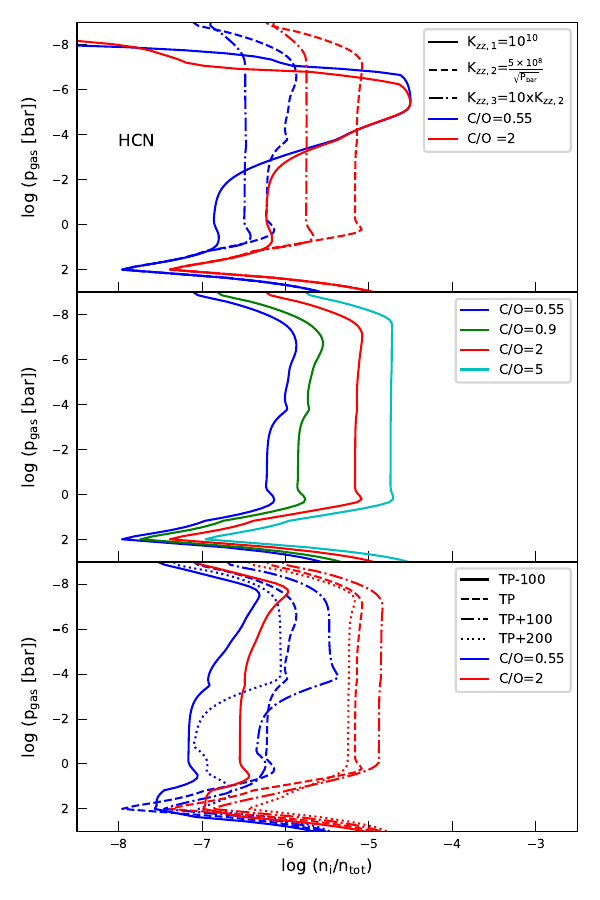}
\caption{Same as Figure \ref{H2O_CO} but for HCN.}
\label{HCN}
\end{figure}

In these models, HCN is generated in the deep atmosphere through scheme \ref{HCN formation}. Figure \ref{HCN} illustrates that for stronger diffusion HCN is quenched at lower concentrations in the deeper regions of the atmosphere. Conversely, less efficient diffusion results in higher quenched HCN concentrations since it is quenched at higher altitudes, where NH$_3$ and CH$_4$ concentrations are also quenched, allowing for local chemical production. In the upper atmosphere, a lower K$_{\rm{zz}}$ supports the concentrations (until it is photochemically destroyed), because NH$_3$ photodissociation is not offset by diffusion.
The higher concentrations of CH$_4$ at larger C/O results in more NH$_3$ chemically converted to HCN for high C/O ratios. In addition, the higher C/O also smooths out the HCN profile throughout the atmosphere as the quenched HCN concentrations approach those produced by photochemistry. 
HCN concentrations initially rise with temperature up until the (T$_{\rm gas}$, p$_{\rm gas}$) profile ``TP+200'', at which profile the concentration decreases. This is because the reaction rate coefficient of HCN formation from CH$_4$ and NH$_3$ increases as temperatures rise, while the concentrations of NH$_3$ and CH$_4$ themselves decrease. HCN is primarily destroyed in these models by being recycled back into NH$_3$ as follows 
\begin{equation}
    \begin{aligned}
        \ce{HCN + M &-> HNC + M}\\
        \ce{HNC + OH &-> HNCO + H}\\
        \ce{HNCO + H &-> CO + NH2}\\
        \ce{NH2 + H2 &-> NH3 + H}\\
        \ce{H2O + H &-> OH + H2}\\
        \hline
        {\rm Net:~} \ce{HCN + H2O &-> CO + NH3}\\
        \hline
        \label{HCN destruction}
    \end{aligned}
\end{equation}

\section{Comparison to observational data: mid-IR transit spectroscopy } \label{sec:fitting}

Of the fourteen models computed in this work, eight models (Table \ref{cc_results}) are found to have cross-correlation maps that peak at the expected location (v$_\mathrm{rest}$$\sim$0 km/s and K$_\mathrm{P}$$\sim$K$_\mathrm{P,theo}$ = 127.11$^{+ 1.49 }_{- 1.52 }$ km/s), where K$_\mathrm{P,theo}$ refers to the expected radial-velocity semi-amplitude of WASP-69b(see Table~\ref{cc_results}). The S/N maps for all models computed in this work are shown in Figure~\ref{appfig}.

The CC framework implicitly assumes that the signal amplitude and the S/N are uniform across the considered spectral range. Yet, this assumption is only valid at a first-order level. To delve deeper into this aspect, we transformed the CC values into likelihood (LH) mapping values for the eight models following the method proposed by \citet{BrogiLine2019} and utilized by \citet{Giacobbe2021}. In this approach, LH is determined by considering the model’s line depth alongside the S/N order by order and spectrum by spectrum. Consequently, identifying a peak in the LH at the correct position within the (v$_\mathrm{rest}$, K$_\mathrm{P}$) space provides additional confirmation of the detection. Yet, the eight models are compatible within 2$\sigma$, suggesting that they are statistically equivalent.

\begin{table*}
\centering
\begin{tabular}{c|c c c | c c c c }
\hline
\hline
model number & \multicolumn{3}{c}{CC framework} & \multicolumn{4}{|c}{LH framework} \\
\hline
                                  &  v$_\mathrm{rest_0}$ [km/s]&  K$_\mathrm{P_0}$ [km/s] & S/N  & v$_\mathrm{rest_0}$ [km/s]& K$_\mathrm{P_0}$ [km/s]& LH$_\mathrm{max}$  & $\sigma$ \\
                                  \hline
     
    1 & 0.0 & 105.0 $^{+ 81.0 }_{- 49.5 }$& 3.18 & 2.0& 138.0& 5527572.94 & 1.86\\ 
    3 & 0.0 & 102.0 $^{+ 73.5 }_{- 48.0 }$& 3.21 & 2.0& 135.0& 5527573.63 & 1.44\\
    4 & 0.0 & 100.5 $^{+ 72.0 }_{- 46.5 }$& 3.04 & 2.0& 138.0& 5527573.66 & 1.42\\
    7 & 0.0 & 105.0 $^{+ 84.0 }_{- 45.0 }$& 3.48 & 2.0& 141.0& 5527573.74 & 1.37\\
    9 & 0.0 & 105.0 $^{+ 91.5 }_{- 48.0 }$& 3.23 & 2.0& 141.0& 5527574.67& Ref.\\
    10 & 0.0 & 102.0 $^{+ 76.5 }_{- 46.5 }$& 3.15& 2.0& 141.0& 5527574.23& 0.94\\ 
    12 & 0.0 & 133.5 $^{+ 66.0 }_{- 70.5 }$& 3.35 & 0.0& 159.0& 5527573.83 & 1.30\\
    13 & 0.0 & 102.0 $^{+ 75.0 }_{- 48.0 }$& 3.21 & 2.0& 138.0& 5527574.00 & 1.16\\
   
      \hline      
\end{tabular}
\caption{Cross-correlation and Likelihood test results}
\tablecomments{From left to right: the model name, the CC result with theoretical models in the S/N framework, and the LH findings. The planet orbital semi-amplitude (K$_\mathrm{P_0}$) as well as the velocity in the planet rest frame (v$_\mathrm{rest_0}$) of the CC and LH peak are reported. The S/N, the maximum of the likelihood matrix LH$_\mathrm{max}$, and the goodness of fit obtained with the Wilk’s theorem on the LH-ratio test are also present. The goodness of fit of the models is shown with respect to the best model in units of standard deviations $\sigma$ (the higher $\sigma$, the more disfavored the model).}
\label{cc_results}
\end{table*}

This study represents one of the first applications of high-resolution data to derive atmospheric properties based on kinetic chemistry modeling. Among the eight models fitting the data in CC, the four exhibit C/O of 2.0, including model 9, which yields the best fit in the likelihood analysis. This aligns with \citet{Guilluy}'s findings of a carbon-rich atmosphere (C/O\,=\,2.0) on WASP-69b. Conversely, the remaining four models suggest C/O of 0.55 and 0.9, preventing a definitive conclusion of a carbon-rich atmosphere. Furthermore, seven of the eight models suggest an atmosphere with a temperature-pressure profile comparable to or cooler than equilibrium modeling predictions at the sub-stellar point, while only one indicates a hotter atmosphere. Finally, the observational data used here does not conclusively constrain the shape or value of the K$_{\rm{zz}}$ parameter.

\begin{figure}
   \centering
   \includegraphics[width = 0.75\linewidth]{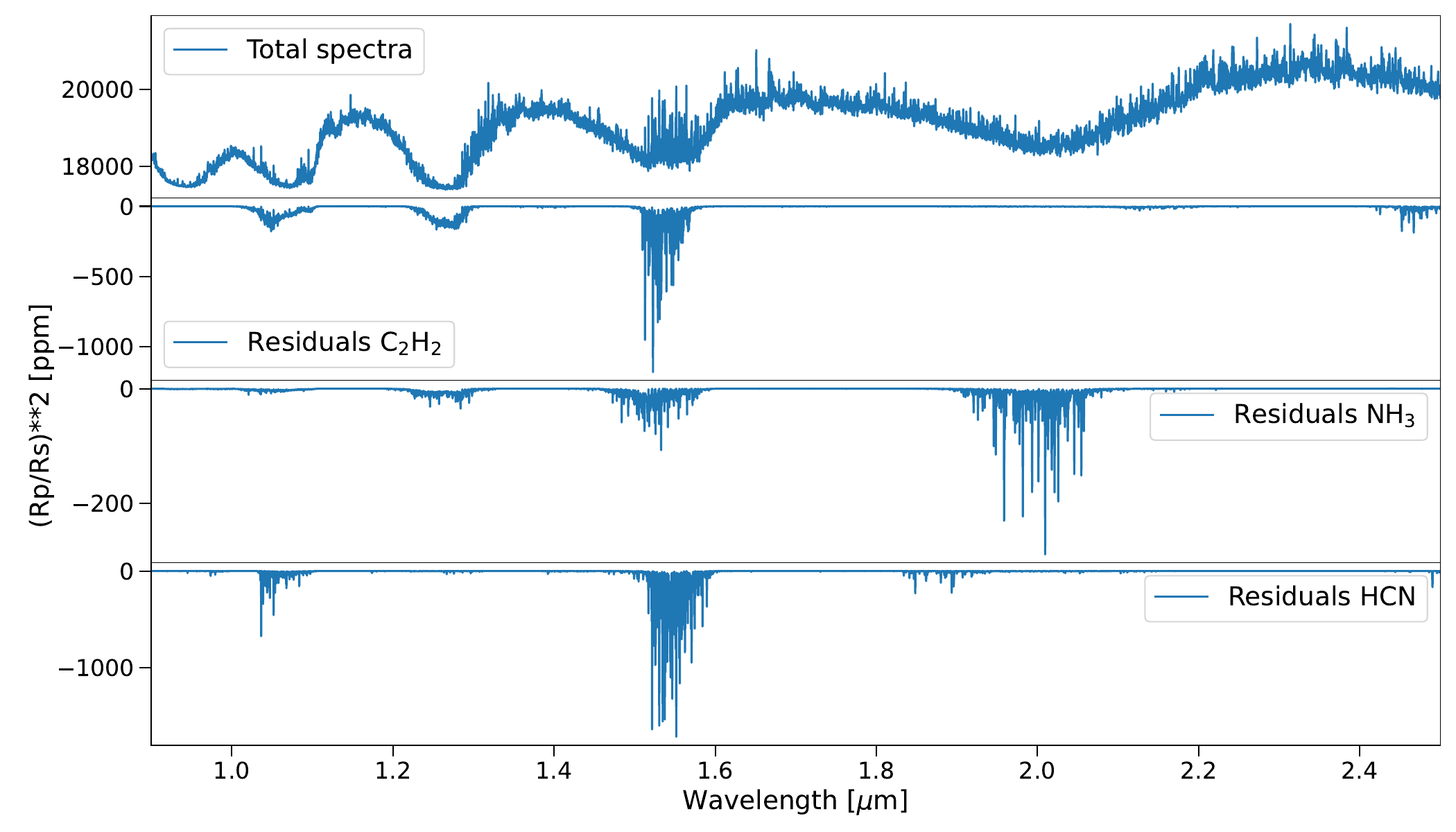}
   \caption{Synthetic transmission spectra for model 12 with indications for the C$_2$H$_2$, HCN and NH$_3$ signal strengths.}
              \label{HCN spectra}%
    \end{figure}

Our results highlight the intrinsic degeneracy in the problem at hand, at least within the scope of our grid of models. One of the possible reasons for this degeneracy is that the physical and chemical properties of planetary atmospheres are not homogeneous across the terminator (see Section \ref{5.1}). One possible additional explanation is that transmission spectra generated with different parameter combinations could appear similar. Finally, we acknowledge that there are data analysis steps yet to be fully understood, especially when high-resolution observations are used to derive atmospheric properties. One such aspect is the order selection used for the CC analysis \citep[][Giacobbe et al., in prep]{Guilluy}. To address these challenges in future works, retrievals could be applied, although this may lead to non-physical solutions. Nonetheless, the computation of physically motivated models is still too slow to be applied in a retrieval framework, and therefore it is crucial to compare the results of retrieval analyses to those obtained from physically motivated forward modeling.

\section{Discussion} \label{sec:discussion}
\subsection{Multi-dimensional nature of planetary atmospheres} \label{5.1}
If the atmosphere of WASP-69b is indeed carbon-rich, as indicated by four out of eight models that match the observations, the temperature variations among these models may reflect the contrasting conditions of the morning (cooler) and evening (warmer) terminators. These differences align with the expected temperature gradients for a planet of this type \citep{Helling2023}. As the observation is time and space-averaged, contributions from both regions can be expected.  Efforts to disentangle the components from the two terminators on exoplanets have been a focus of several recent studies, utilizing both low-resolution and high-resolution data to reveal differing atmospheric conditions across the two regions \citep[e.g.][]{Ehrenreich2020,Espinoza2021, Grant&Wakeford2023}.

A notable observation is the detection of C$_2$H$_2$ in WASP-69b's atmosphere, accompanied by the lack of detection of HCN. Both molecules are commonly considered tracers of warm, carbon-rich atmospheres \citep{Venot2015, Ohno2022}. Among our best fit models, only model 12 - which features the highest temperature within the carbon-rich models - shows a strong C$_2$H$_2$ signal. This model also predicts a substantial HCN concentration with a spectral signal comparable to that of C$_2$H$_2$, and over ten times stronger than that of NH$_3$, particularly near 1.5$\mu$m (see Figure \ref{HCN spectra}). Therefore, if C$_2$H$_2$ is present on WASP-69b, we predict that HCN should also exist at detectable levels, suggesting that future observations might reveal its signature.

Our 1D modeling does not account for 3D mixing effects and wind jets. In tidally locked hot Jupiters, uneven heating may result in significant temperature contrasts between the day and night sides, leading to variations in atmospheric composition. Strong super-rotating equatorial jets may instead homogenize these differences in composition \citep[e.g.][]{Madhusudhan2016,Baeyens2022a}. In line with the results of \cite{Baeyens2023}, it can be anticipated that the main effect of horizontal chemical transport in WASP-69b would be to quench molecular concentrations on the evening terminator to values typical of the hotter day-side regions and those on the morning terminator to levels resembling the colder night side. Given that the equilibrium temperature of WASP-69b $\sim$ 950 K puts it on the cusp of CH$_4$-CO conversion and NH$_3$-N$_2$ conversion, and that our models show that temperature differences of 100-200 K through the atmosphere significantly changes the concentrations of the observed molecules, this horizontal quenching may result in significantly different concentration profiles for the two terminator regions. An additional consequence of the horizontal mixing would be an increase in photochemical products HCN and C$_2$H$_2$ on the morning terminator due to mixing of the more photochemical active parent species CH$_4$ and NH$_3$ from the night-side, while there may be fewer of these photochemical products on the evening terminator, as demonstrated by \citet{Tsai2024}. 

\subsection{Capturing NH$_3$ Quenching} \label{5.2}
\begin{figure*}
\centering
\includegraphics[width = 0.8\linewidth]{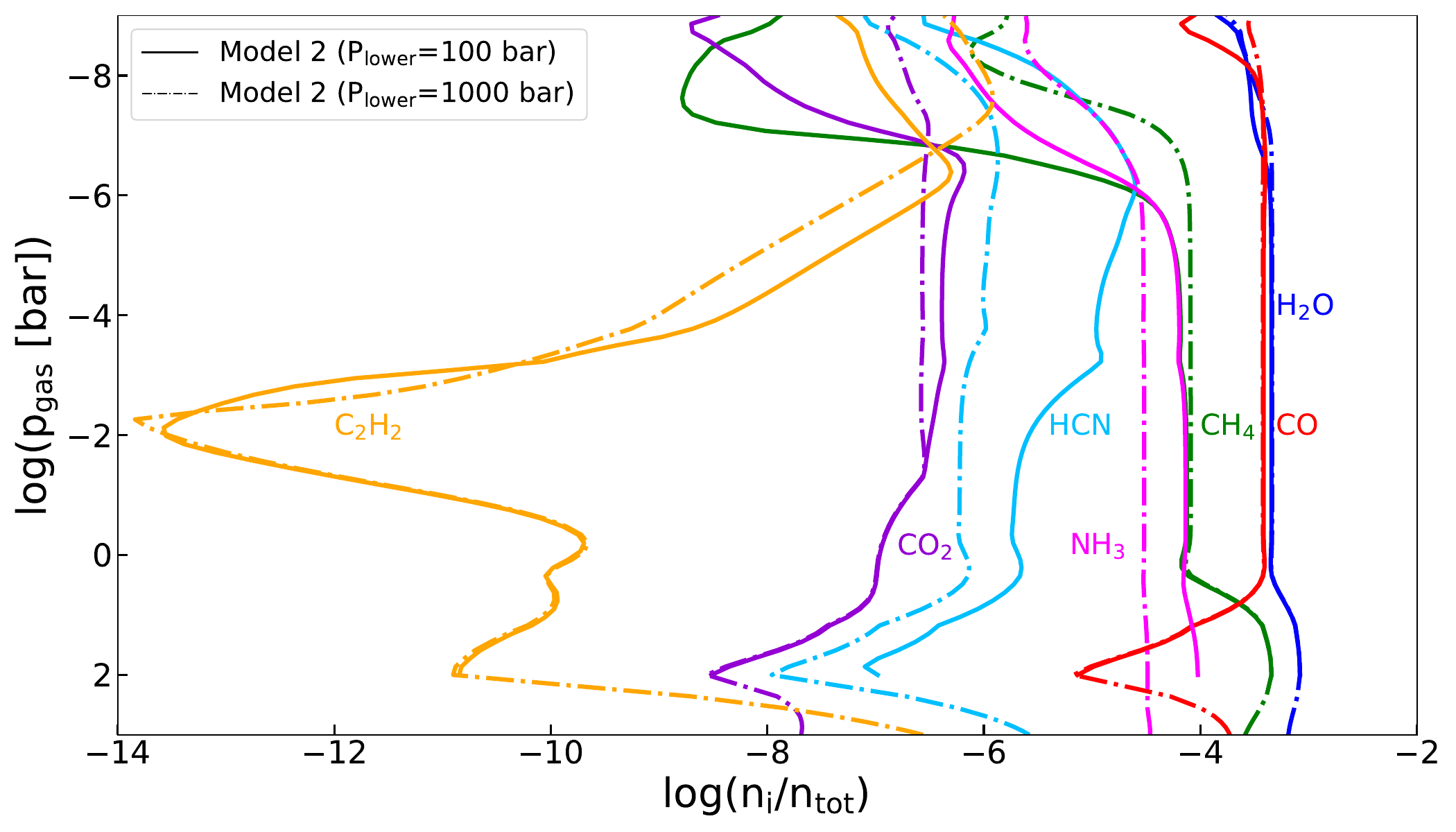}
\caption{Concentrations of key species for the terminator region of WASP-69b, for a model with lower boundary 100 bars vs lower boundary 1000 bars.} 
\label{deep atmo}
\end{figure*}
The (T$_{\rm gas}$, p$_{\rm gas}$) profile retrieved through equilibrium modeling in \citet{Guilluy} is constrained to a maximum pressure of 100 bars. To evaluate the implications of this constraint on non-equilibrium chemistry, we compare our models with lower boundary pressures of 1000 bars to those restricted to 100 bars. 

Our models are initialized with atomic abundances at the lower boundary, as opposed to equilibrium values, and are then evolved until a steady state is achieved. This approach reveals significant deviations from equilibrium concentrations for NH$_3$ and HCN at 100 bars (Figure \ref{deep atmo}), indicating that NH$_3$ quenching occurs at pressures deeper than 100 bars, in agreement with the findings of \citet{Moses2011} for HD189733b, where NH$_3$ concentrations at 100 bars are strongly influenced by the N$_2$-NH$_3$ interconversion quench point at higher temperatures and pressures.

At lower atmospheric temperatures, the kinetics pathways favour NH$_3$ formation. For WASP-69b, the deeper atmospheric layers are cooler than that of HD189733b. As a result, the 100 bar models over-predict NH$_3$ concentrations relative to their equilibrium value, and produce higher NH$_3$ concentrations than expected for the deeper pressure models. This overestimation also impacts the formation of HCN, amplifying its abundance. In contrast, other key molecules such as CH$_4$, CO, and H$_2$O show minimal sensitivity to the pressure boundary, as their concentrations are largely dictated by equilibrium chemistry even at 100 bars, and they are quenched at lower pressures, minimizing their dependence on deeper atmospheric dynamics.

To evaluate the observational implications of these differences, we generated synthetic spectra for the 100-bar models. The spectra consistently show stronger HCN spectral features compared to NH$_3$, reflecting the higher HCN concentrations across these models. Additionally, only three of the fourteen 100-bar models achieve robust CC fits and likelihoods, compared to the eight models in the 1000-bar atmosphere models. The three good fit models in this case correspond to the parameters of Model 7, 9, and 12 of table \ref{parameters}, thus biasing the conclusions to carbon-rich atmospheres (C/O=2). This difference highlights the importance of accurately capturing the NH$_3$
quenching at high pressures.

Uncertainty in individual rate coefficients is another factor that may influence the quenched NH$_3$ concentrations in high pressure and temperature environments. The N$_2$-NH$_3$ interconversion is particularly sensitive to non-equilibrium chemistry, and differences of over an order of magnitude in the relevant reaction rate coefficients are reported in literature \citep[e.g.][]{Moses2014,Tsai2021,Veillet2024}. For example, \citet{Moses2014} demonstrates that NH$_3$ predictions for HD189733b could differ by up to an order of magnitude depending on whether the \citet{Moses2011} or \citet{Venot2012} chemical network was employed. \citet{Rimmer&Helling} provides a further discussion on the choice of reaction rate coefficients for NH$_3$ adopted in STAND, and a comparison with the \citet{Moses2011} network.


\section{Summary} \label{sec:conclusions}
A parameter study of K$_{\rm{zz}}$, (T$_{\rm gas}$, p$_{\rm gas}$) and C/O ratio was conducted with 1D photochemistry-diffusion modeling to analyse the effect of vertical mixing and photochemistry in producing the species detected in WASP-69b's atmosphere, and further constrain C/O. 
With this study, we confirm the hypothesis of Guilluy et al. (2022) that the near-IR WASP-69b transition spectrum is more likely shaped by chemical disequilibrium processes rather than equilibrium; however, our exploration of 1D disequilibrium chemistry prescriptions cannot replicate all of the spectral features observed on WASP-69b.
Our results indicate that 
   \begin{enumerate}
      \item CH$_4$, NH$_3$, C$_2$H$_2$ and HCN are the species most sensitive to vertical diffusion and photochemistry in WASP-69b. Of the three parameters tested in this work, the local gas temperature has the strongest impact on the concentrations of these species in the terminator regions. However, this dependency cannot be easily disentangled from that of atmospheric C/O. 
      \item A highly efficient mixing in the upper atmosphere competes with the effect of photodissociation, effectively retaining the photochemical parent species at their quenched concentrations and thereby reducing the production of species like C$_2$H$_2$ and HCN.
      \item A combination of high gas temperatures and C/O $\sim$ 2 is required to produce C$_2$H$_2$ concentrations at the 10$^{-7}$--10$^{-6}$ level in the lower atmosphere. However, at gas pressures $<10^{-6}$ bars, high concentrations of C$_2$H$_2$ are also photochemically produced for solar C/O.
      \item The relation of C$_2$H$_2$ concentrations and K$_{\rm{zz}}$ is complex and dependent on the local gas temperature and composition (i.e. C/O). 
      \item The observational data is not sufficient to distinguish different mixing profiles.
   \end{enumerate}
\noindent

Metallicity has not been examined in this study as a free parameter. However, if atmospheric metallicity were to increase, the resulting reduction in CH$_4$ concentrations would cause a corresponding decrease in HCN and C$_2$H$_2$ concentrations, due to the balance between CO and CH$_4$ production shifting towards CO \citep{Soni2023}. \citep{Moses2013} find that although the CH$_4$/H$_2$O fraction is more sensitive to C/O than to metallicity, the individual concentrations are very sensitive to metallicity. In terms of NH$_3$, \cite{Ohno2023} found that although the quenched NH$_3$ concentration remains constant as atmospheric metallicity increases, the ratio of NH$_3$ to bulk atmospheric nitrogen abundance decreases significantly, as N becomes locked in N$_2$. Therefore, higher metallicity can be expected to result in lower concentrations of CH$_4$, HCN, C$_2$H$_2$ and NH$_3$. As outlined in \cite{Guilluy}, this scenario is disfavored by the observations. 

After performing cross-correlation of the simulated transmission spectra for the fourteen models with the observed high-resolution data, eight models were identified with the expected planetary rest-frame velocity v$_\mathrm{rest_0}$ = 0 km/s and planetary maximal radial-velocity K$_\mathrm{P}$$\sim$K$_\mathrm{P,theo}$. These models all differ in C/O and K$_{\rm{zz}}$, but suggest a temperature-pressure profile consistent with the predictions of equilibrium modeling, or cooler. The difficulty in distinguishing between models of varying (T$_{\rm gas}$, p$_{\rm gas}$) and K$_{\rm{zz}}$ might be related to the spatial and temporal averaging of the observations. Interestingly, despite the high C/O models suggesting high HCN concentrations, this molecule has not been detected in the atmosphere of WASP-69b. The transmission and emission spectroscopy observations of WASP-69b with JWST (GTO programs 1177 and 1185, and GO program 3712) can be expected to provide more insights on its atmospheric properties. By combining the low- and high-resolution observations, the atmospheric properties of this planet may be better constrained and a better understanding obtained of the kinetic processes in its deep and it upper atmosphere.
 
\begin{acknowledgments}
N.B. acknowledges financial support from the Austrian Academy of Sciences. Ch.H. is part of the CHAMELEON MC ITN EJD which received funding from the European Union’s Horizon 2020 research and innovation programme under the Marie Sklodowska-Curie grant agreement number 860470. P.C. is funded by the Austrian Science Fund (FWF) Erwin Schroedinger Fellowship, program J4595-N. G.G. and P.G acknowledge financial contribution from PRIN MUR 2022 (project No.\,2022CERJ49) and PRIN INAF 2019. N.B would like to thank Patrick Barth for helpful advice regarding use of the ARGO code and STAND network.
\end{acknowledgments}

\vspace{5mm}

\bibliography{biblio}{}
\bibliographystyle{aasjournal}

\begin{appendix}

\section{Additional Figures and Tables}\label{Appendix A}

\begin{figure}[ht!]
   \centering
   \includegraphics[width = 0.5\linewidth]{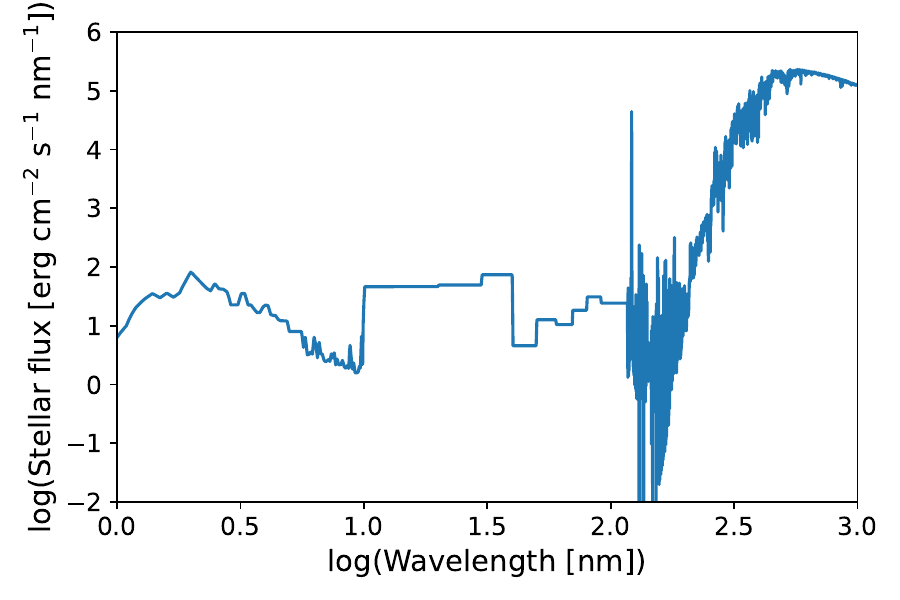}
   \caption{Spectral energy distribution (SED) of the host star at top of the atmosphere of WASP-69b. The SED is constructed by mixing a PHOENIX model in the UV, optical and infrared \citep{Husser2013} and a scaled solar flux in the EUV and X-ray regimes \citep{Claire2012}.}
              \label{FigGam}%
    \end{figure}

\begin{figure}
\centering
\includegraphics[width=\linewidth]{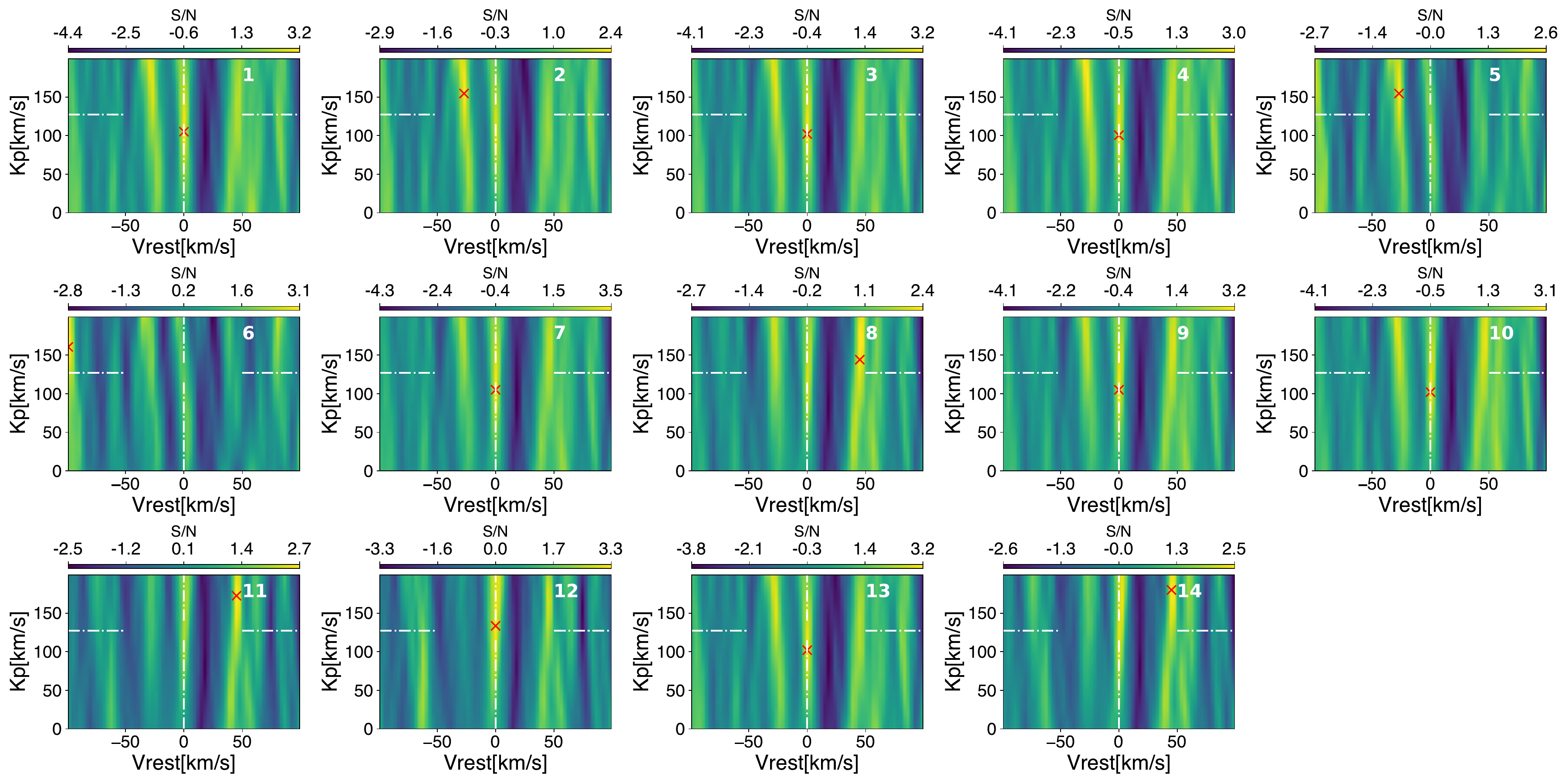}
\caption{Cross correlation in S/N of all models produced in this work to the high-resolution transmission spectroscopy of WASP-69b conducted by \citet{Guilluy} as a function of the planet’s maximum radial velocity (KP) and the planet’s rest-frame velocity (Vrest). The numbers inset correspond to the models listed in Table \ref{table:1}. The CC peaks are indicated by a red cross.}
\label{appfig}
\end{figure}

\begin{figure}
\centering
\includegraphics[width=0.75\linewidth]{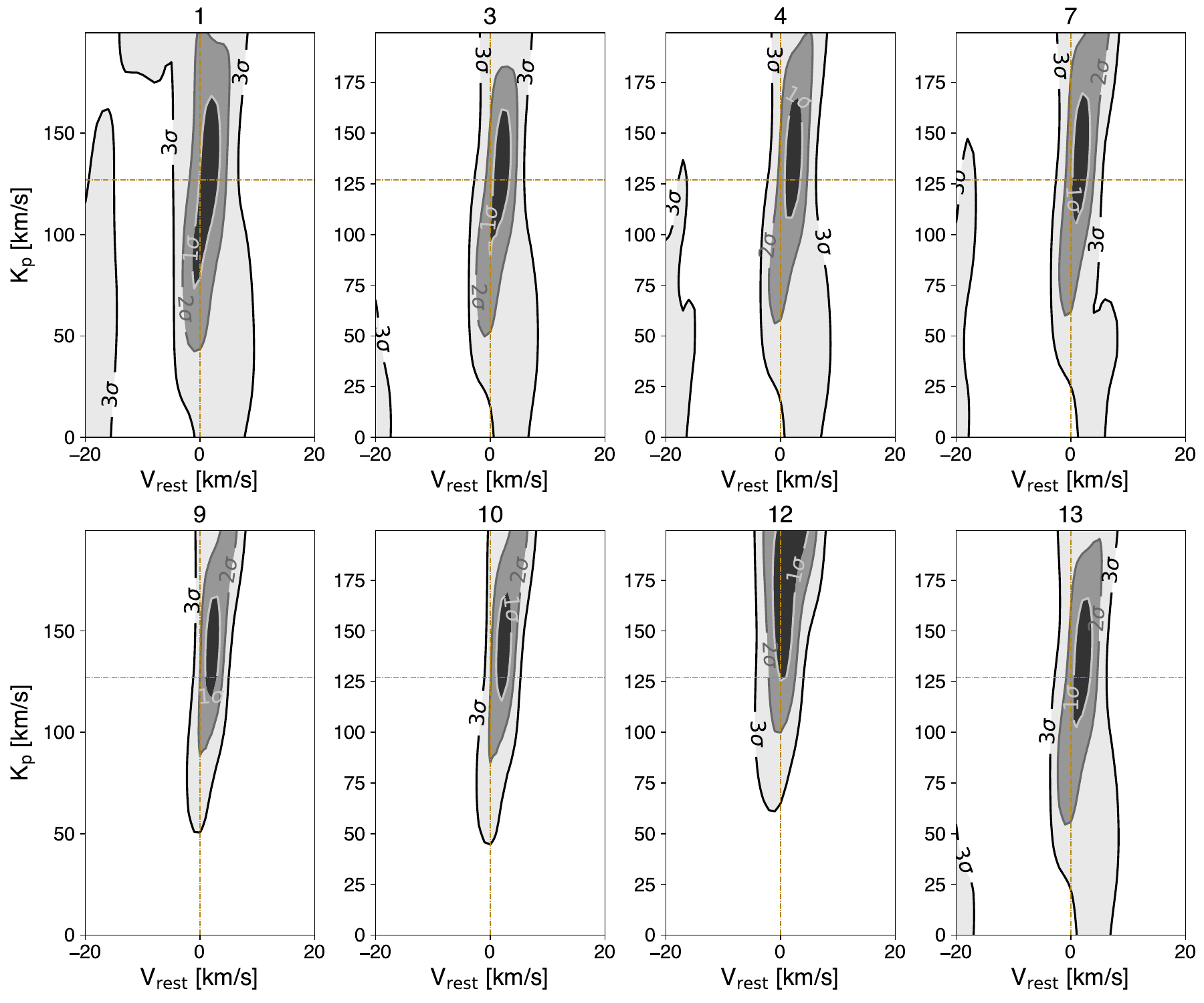}
\caption{LH confidence intervals maps for the models with v$_\mathrm{rest_0}$ = 0 km/s and K$_\mathrm{P}$$\sim$K$_\mathrm{P,theo}$ in CC. The plot titles correspond to the model descriptions in Table \ref{table:1}.}
\label{appfig2}
\end{figure}

Figure \ref{FigGam} presents the input stellar flux at the top of the atmosphere of WASP-69b. Figure \ref{appfig} shows the posterior distribution for the S/N fitting of disequilibrium model spectra produced in this work to the high-res data obtained by \citet{Guilluy}. In Figure \ref{appfig2}, the LH maps for the eight models with v$_\mathrm{rest_0}$ = 0 km/s and K$_\mathrm{P}$$\sim$K$_\mathrm{P,theo}$ are shown. 

\end{appendix}

\end{document}